\documentclass[%
reprint,          
  ％groupedaddress,    
  nofootinbib,       
  amsmath,amssymb,
  aps,
  prd,
  longbibliography,
]{revtex4-2}

\usepackage{graphicx}
\usepackage{dcolumn}
\usepackage{bm}

\usepackage[colorlinks=true, linkcolor=blue, citecolor=blue, urlcolor=blue]{hyperref}

\numberwithin{equation}{section}
\renewcommand{\theequation}{\arabic{section}.\arabic{equation}}

\begin{document}

\preprint{APS/123-QED}

\title{Skewness in the Hellings-Downs curve}


\author{Ryosuke Fujimoto}
\email{MTryou369@gmail.com}
\author{Keitaro Takahashi}%
\affiliation{%
Faculty of Science, Kumamoto University, Japan
}%

\date{\today}

\begin{abstract}
Recent Pulsar Timing Array datasets provide compelling evidence for a nano-Hertz gravitational-wave background, but robust detection requires characterizing statistical fluctuations of the Hellings–Downs (HD) correlation expected from a finite population of discrete sources. Building on the variance calculation of Allen (2023), we derive the third central moment (skewness) of the HD correlation for a single unpolarized point source and an ensemble of many interfering point sources in the confusion-noise regime. To isolate the intrinsic non-Gaussianity of the background, we extend the pulsar-averaging formalism to third order by introducing a three-point averaged correlation function, which allows us to define the cosmic skewness. We find that the skewness remains non-zero in the large-source-number limit and is controlled by a new geometric three-point function. These results suggest that incorporating higher-order moments could provide additional information on source discreteness beyond standard Gaussian analyses.
\end{abstract}

\maketitle


\section{\label{sec:Introduction}introduction}
With the advancement of gravitational wave (GW) astronomy, Pulsar Timing Arrays (PTAs) have gathered significant attention as a method for detecting GWs in the nanohertz (nHz) frequency band~\cite{Detweiler1979, FosterBacker1990, BurkeSpolaor2019}. Functioning as a galactic-scale detector, a PTA utilizes a network of millisecond pulsars distributed throughout the Milky Way~\cite{FosterBacker1990, IPTA2016_DR1}. By precisely measuring the times of arrival (TOAs) of pulses from these pulsars, observers on Earth can detect minute fluctuations caused by distortions in spacetime~\cite{Sazhin1978, Detweiler1979}.

Indeed, recent observations indicate that the field is at a historic turning point. In 2023, research teams worldwide---including NANOGrav (North America)~\cite{Agazie:2023,NANOGrav_2024_Methods}, EPTA-InPTA (Europe and India)~\cite{Antoniadis:2023}, PPTA (Australia)~\cite{Reardon:2023}, and CPTA (China)~\cite{Xu:2023}---successively announced compelling evidence for the existence of a GW background in the nanohertz frequency band, based on years of accumulated observational data.

However, isolating and detecting a single GW source distinguishable from noise remains challenging for current PTA observations~\cite{Ravi2012, Babak2013, Shannon2015}. Consequently, PTA groups currently focus on detecting a stochastic gravitational wave background (SGWB), which arises as a superposition of multiple unresolved signals~\cite{Phinney2001, JaffeBacker2003, Sesana2008, Sesana2013, Rosado2015,Christensen:2018}.

The definitive signature of such an isotropic SGWB is a specific spatial correlation in the timing residuals between pulsar pairs, known as the Hellings-Downs (HD) correlation~\cite{HellingsDowns1983}. Predicted by Hellings and Downs (1983), this curve describes the expected correlation as a function of the angular separation $\gamma$ between pulsars, serving as the "smoking gun" evidence that distinguishes a true GW background from other noise sources~\cite{HellingsDowns1983, Jenet2005}.

While the standard HD correlation represents the ensemble average (mean) of these correlations, realistic observations involving a finite number of sources exhibit statistical fluctuations~\cite{Roebber2016, 2022JCAP...11..046B}. Recently, Allen (2023) provided a rigorous analytical derivation of these fluctuations, calculating the variance of the HD correlation for an ensemble of many point sources~\cite{Allen2023_Variance, AllenRomano2023}. They introduced an observationally motivated procedure to decompose the total variance into a reducible pulsar-variance component and an irreducible cosmic-variance component by pulsar-averaging before forming moments. Because cosmic variance cannot be removed by increasing the number of pulsars, the variance itself becomes an additional observable that can carry information about the nature of the GW sources beyond the mean HD curve.

In this paper, we proceed with our discussion based on the theoretical framework of the variance calculations established by Allen (2023), extending the analysis to higher-order statistical properties. As Allen (2023) emphasized, even if the timing residuals are Gaussian, the induced inter-pulsar correlations are not Gaussian and exhibit intrinsic non-Gaussianity. For a stochastic background generated by a finite number of discrete GW sources, additional non-Gaussian features associated with source discreteness are expected. Such non-Gaussianity cannot be fully captured by the mean HD correlation or the variance alone; it is instead characterized by higher-order moments such as the skewness~\cite{2025JCAP...01..017B}. Motivated by this, we take a first step toward quantifying non-Gaussianity in PTA correlations by evaluating the skewness of the inter-pulsar correlation distribution. This provides a natural extension of variance-based analyses and offers complementary information for interpreting departures from the HD mean.

The remainder of this paper is organized as follows. In Sec.~\ref{sec:single}, we establish the foundational framework by defining the antenna pattern functions and the unpolarized correlation for a single GW source. Following the methodology of Allen (2023)~\cite{Allen2023_Variance}, we review the calculation of the mean and variance, and then explicitly derive the skewness for the single-source case.
In Sec.~\ref{sec:many}, we extend the analysis to the "confusion noise limit" involving many interfering point sources. For the skewness in this interfering model, we specifically focus on the dominant terms that characterize the primary non-Gaussian signature of the background.
In Sec.~\ref{sec:cosmic}, we decompose the total fluctuations into pulsar variance and cosmic variance. To evaluate the higher-order statistical nature of the SGWB itself, we define the "three-point-average function" of the HD correlation and use it to derive the cosmic skewness.
Sec.~\ref{sec:Discussion and Summary} provides the conclusions of this study and a discussion on the potential use of cosmic variance/skewness to improve GWB analyses. Detailed calculations of the many-point skewness are presented in Appendix~\ref{app:many}, and we provide the basic equation to numerically compute the three-point-average function in Appendix~\ref{app:cosmic}.

\section{\label{sec:single}Single GW Source Statistics}

In this section, we define and review the basic components and statistical quantities associated with the HD curve for a single point source~\cite{Allen2023_Variance}.
In the correlation analysis of GW signals using PTAs, antenna pattern functions are fundamental quantities defined for each pulsar and polarization mode~\cite{HellingsDowns1983, Anholm2009}. The antenna pattern function \(F_\alpha^A(\boldsymbol{\Omega})\) for a pulsar located at direction \(\boldsymbol{p}_\alpha\) and a GW coming from direction \(\boldsymbol{\Omega}\) is given by~\cite{Detweiler1979, Allen2023_Variance}:
\begin{align}\label{eq:antenna}
F^A_\alpha (\boldsymbol{\Omega})
  = \frac{1}{2} 
  \frac{\boldsymbol{p}^a_\alpha \boldsymbol{p}^b_\alpha}{1 + \boldsymbol{\Omega} \cdot \boldsymbol{p}_\alpha} 
  e_{ab}^A(\boldsymbol{\Omega}),
\end{align}
where \(e_{ab}^A(\boldsymbol{\Omega})\) is the polarization tensor corresponding to the GW direction \(\boldsymbol{\Omega}\), and the index \(A\in \{+,\times\}\) denotes the GW polarization modes (plus and cross)~\cite{Maggiore2000}.
The unpolarized correlation \(\rho_\text{u}\) generated by a single unpolarized point source at \(\boldsymbol{\Omega}\) between the pulsar pair \(\boldsymbol{p}_1,\,\boldsymbol{p}_2\) is described using the product of these antenna pattern functions~\cite{CornishSesana2013, Allen2023_Variance}:
\begin{align}\label{eq:single integrand}
\rho_\text{u}(\boldsymbol{\Omega}) 
 =F_1^+(\boldsymbol{\Omega})F_2^+(\boldsymbol{\Omega}) 
+ F_1^{\times}(\boldsymbol{\Omega}) F_2^{\times}(\boldsymbol{\Omega}).
\end{align}

\subsection{Mean of the Single Source Correlation}\label{sub:single mean}
The statistical properties of this single-source correlation \(\rho(\boldsymbol{\Omega})\) are determined by averaging over the source direction \(\boldsymbol{\Omega}\) at a fixed pulsar angular separation \(\gamma = \arccos(\boldsymbol{p}_1\cdot\boldsymbol{p}_2)\)~\cite{Allen2023_Variance}. This averaging process, denoted by \(\langle...\rangle_{\boldsymbol{\Omega}}\equiv\frac{1}{4\pi} \int \mathrm{d}\boldsymbol{\Omega}(...)\), yields the mean, variance, and skewness of the correlation~\cite{Allen2023_Variance, Roebber2016}.

The mean of the correlation corresponds to the well-known HD curve \(\mu_\text{u}(\gamma)\)~\cite{HellingsDowns1983}:
\begin{align}\label{eq:single mean mu}
  \mu_\text{u}(\gamma) 
  &=\langle\rho_\text{u}\rangle_{\boldsymbol{\Omega}}\notag\\
  &=\frac{1}{4} + \frac{1}{12} \cos\gamma + 
   \frac{1}{2}(1-\cos\gamma)\ln\left(\frac{1-\cos\gamma}{2}\right).
\end{align}
This function represents the expected correlation for an isotropic, unpolarized, and continuous stochastic gravitational wave background~\cite{HellingsDowns1983, Jenet2005, RomanoCornish2017}.

\subsection{\label{sub:single var}Single Source Variance}
Next, we define the variance functions for a single source, which are required for the many-point statistics discussed in Sec.~\ref{sec:many}. These functions are derived by taking the second moment of the antenna pattern combinations over the source direction \(\boldsymbol{\Omega}\)~\cite{Allen2023_Variance}. Detailed derivations are provided by Allen (2023)~\cite{Allen2023_Variance}.

\begin{enumerate}
  \item \textbf{Unpolarized Variance} \((\sigma^2_\text{u})\): This represents the standard variance of the HD curve. It arises from the "unpolarized" correlation term \(\rho_\text{u} = F_1^+F_2^+ + F_1^{\times} F_2^{\times}\) and quantifies the scattering around the mean \(\mu_\text{u}\). The analytic expression derived by Allen (2023) is~\cite{Allen2023_Variance}:
\begin{align}\label{eq:unpolarized var}
 \sigma_{\text{u}}^2(\gamma) 
 =& 
 \langle
 \rho_\text{u}^2
 \rangle_{\boldsymbol{\Omega}} -\mu_\text{u}^2\notag\\
 =& \frac{97}{80} + \frac{1}{24} c - \frac{839}{720} c^2 \notag\\
 &+ \frac{1}{12} \Psi(c) \Bigl(18 - 10c - 3\Psi(c)\Bigr),
\end{align}
where 
\begin{align}\label{eq:Psi}
  c &\equiv \cos\gamma, \quad \text{and} \notag\\
  \Psi(c) &\equiv (1 - c)\ln\left( \frac{1 - c}{2} \right).
\end{align}

  \item \textbf{Polarized Variance} \((\sigma^2_\text{p})\): This function arises from the "polarized" correlation term \(\rho_\text{p} = F_1^+F_2^\times - F_1^\times F_2^+\). Since this term has a mean of zero (\(\langle\rho_\text{p} \rangle_{\boldsymbol{\Omega}} = 0\)), its variance is equal to its second moment. The analytical expression is given by~\cite{Allen2023_Variance}:
 \begin{align}\label{eq:polarized var}
   \sigma^2_\text{p}(\gamma) 
    =& \langle \rho_\text{p}^2 \rangle_{\boldsymbol{\Omega}} \notag\\
   =&\frac{7}{6} \left( c^2-1 \right) \notag\\
    &+ \frac{1}{4} (3c-7) (1-c) \ln \left( \frac{1-c}{2} \right).
 \end{align}

  \item \textbf{Cross-Polarized Variance} \((\sigma^2_\text{c})\): This function represents the averaged product of the autocorrelations of each pulsar. It is defined as~\cite{Allen2023_Variance}:
 \begin{align}\label{eq:cross var}
   \sigma^2_\text{c}(\gamma) 
   &= \langle (F_1^+F_1^+ + F_1^{\times} F_1^{\times}) (F_2^+F_2^+ + F_2^{\times} F_2^{\times}) \rangle_{\boldsymbol{\Omega}} \notag\\
   &= \frac{13}{120}+\frac{1}{12}c+\frac{1}{120}c^2.
 \end{align}
 These functions are related by the identity \(\sigma^2_\text{c} =\mu^2_\text{u}+\sigma^2_\text{u} +\sigma^2_\text{p}\), as proven in Appendix F of Allen (2023)~\cite{Allen2023_Variance}.
\end{enumerate}

\subsection{\label{sub:single skew}Single Source Skewness}
The third-order moment, or skewness \(\mathcal{S}_\text{u}(\gamma)\), is the primary focus of this study. It measures the asymmetry of the correlation distribution, serving as a probe for non-Gaussianities in the background~\cite{Bartolo2018}. The skewness is defined using the third central moment \(\kappa_{3,\text{u}}\):
\begin{align}\label{eq:single skewness}
 \mathcal{S}_\text{u}(\gamma)
 &\equiv 
   \frac{\kappa_{3,\text{u}}}{\sigma_\text{u}^3} 
 = \frac{\langle (\rho_\text{u} - \mu_\text{u})^3 \rangle_{\boldsymbol{\Omega}}}{\sigma_\text{u}^3} \notag\\
 &= \frac{\langle \rho_\text{u}^3 \rangle_{\boldsymbol{\Omega}} - 
   3\mu_\text{u} \langle \rho_\text{u}^2 \rangle_{\boldsymbol{\Omega}} 
 + 2\mu_\text{u}^3}{\sigma_\text{u}^3}.
\end{align}
Following the integration techniques outlined by Allen (2023)~\cite{Allen2023_Variance}, we derive the analytical expression for the unpolarized third moment \(\langle \rho_\text{u}^3 \rangle_{\boldsymbol{\Omega}}\) for a single source as follows:
\begin{align}\label{eq:single 3cumulant}
 \langle \rho_\text{u}^3 \rangle_{\boldsymbol{\Omega}} 
 =& \frac{2913}{1120} c^3 - \frac{777}{160} c^2 - \frac{2837}{1120} c 
 + \frac{789}{160}\notag\\
 &+ \frac{3}{8} (3c^2 - 18c + 19) \Psi(c),
\end{align}
where \(c=\cos\gamma\) and \(\Psi(c)\) follows the definition in Eq.~\eqref{eq:Psi}.
Substituting this result into the definition in Eq.~\eqref{eq:single skewness}, along with Eqs.~\eqref{eq:single mean mu} and \eqref{eq:unpolarized var}, yields the unpolarized single-source skewness \(\mathcal{S}_\text{u}(\gamma)\). In the analysis of the third moment, we do not require the polarized and cross-polarized correlations discussed in Sec.~\ref{sub:many skew}, as we present only the dominant term in the many point skewness.

\begin{figure}[t]
  \centering
  \includegraphics[width=0.45\textwidth]{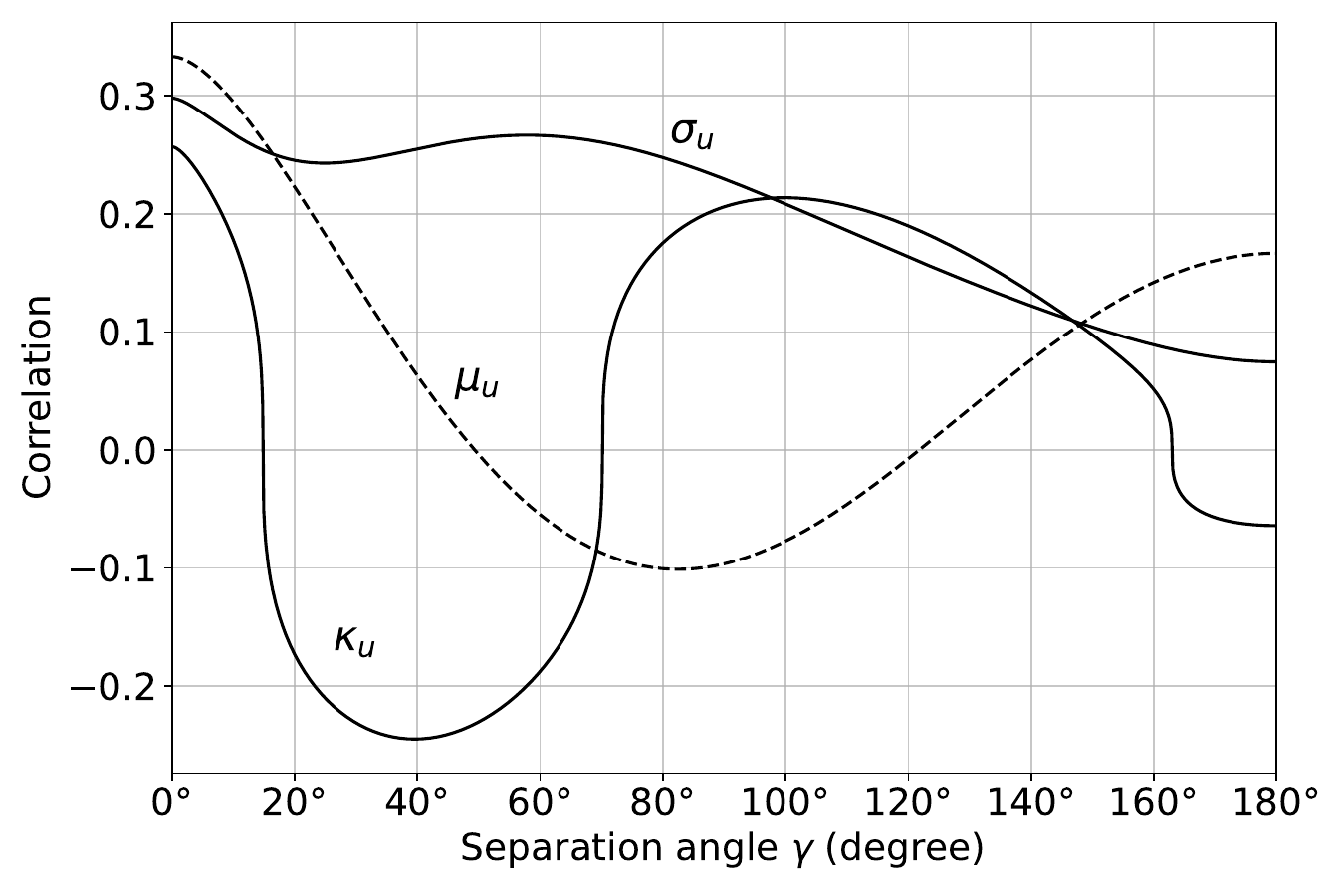}
  \caption{Comparison of single-source statistical quantities. The dashed line represents the HD curve \(\mu_\text{u}(\gamma)\). The solid lines show the unpolarized single-source standard deviation \(\sigma_\text{u}\) and the cubic root of the unpolarized single source skewness numerator \(\kappa_\text{u}\).}
  \label{fig:single_mu_sigma_kappa}
\end{figure}

\begin{figure}[t]
  \centering
  \includegraphics[width=0.45\textwidth]{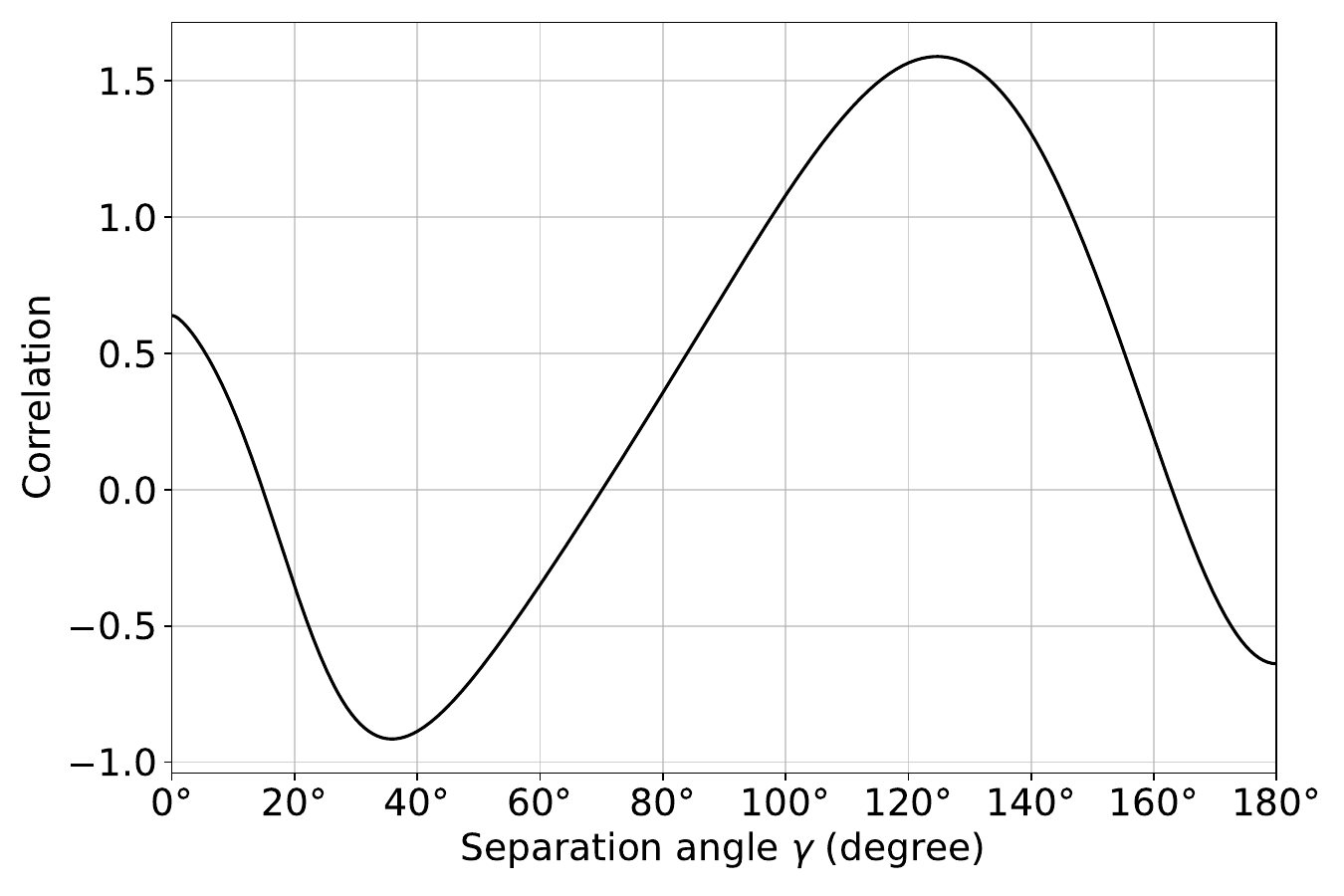}
  \caption{The unpolarized single-source skewness \(\mathcal{S}_\text{u}(\gamma)\) as a function of the angular separation \(\gamma\).}
  \label{fig:single_norm_skew}
\end{figure}

For visualization purposes, we define the cubic root of the skewness numerator as follows: 
\begin{align}\label{eq:root kappa}
  \kappa_\text{u}=\sqrt[3]{\kappa_{3,\text{u}}}.
\end{align}
We plot the unpolarized single source mean, the unpolarized single source standard deviation, and the cubic root of the unpolarized singles source skewness numerator in Fig.~\ref{fig:single_mu_sigma_kappa}. Additionally, the unpolarized single source skewness is presented in Fig.~\ref{fig:single_norm_skew}. These single source statistics are essential for building the many-point statistics discussed in the following sections.

\section{\label{sec:many}Many-Point Statistics of Interfering Sources}
In this section, we analyze the statistical properties of the HD curve observed as a superposition of a finite number \(N\) of discrete GW sources. The analysis focuses on the "interfering source model," where sources radiate at the same angular frequency \(\omega\).
This model is constructed following the "confusion-noise case" discussion in Allen (2023)~\cite{Allen2023_Variance}---a regime where the distinction between a resolvable single source and an unresolved SGWB depends on the signal density and the observation time. This boundary has been classically discussed in Sesana et al. (2009)~\cite{Sesana:2009} and recently explored using efficient simulation methods for realistic populations by Bécsy et al. (2022)~\cite{Becsy:2022}. The spatial distribution of GW sources is assumed to be uniform throughout three-dimensional space~\cite{Phinney2001, JaffeBacker2003}.

Based on this spatial distribution assumption, when the amplitude of the source closest to Earth is denoted as \(\mathcal{A}\), the observed amplitude \(\mathcal{A}_j\) of each source, numbered \(j = 1,2,...,N\) in order of proximity to Earth, is given by the following deterministic formula~\cite{Allen2023_Variance}:
\begin{align}\label{eq:amplitude}
  \mathcal{A}_j = j^{-1/3}\mathcal{A}.
\end{align}
While this amplitude \(\mathcal{A}_j\) can be treated as a probability distribution, this study adopts this deterministic form to enable analytic discussion. We define the sums of powers of the amplitudes, denoted as \(\mathcal{H}_{2n}\):
\begin{align}\label{eq:even amp}
  \mathcal{H}_{2n} 
= \sum_{j=1}^N \mathcal{A}_j^{2n}.
\end{align}
It is important to note that we assume a finite number of sources, \(N\), to avoid Olbers' paradox~\cite{Allen2023_Variance}.

For the calculations in the following subsections, we will specifically require the second, fourth, and sixth-order sums. Following Allen (2023)~\cite{Allen2023_Variance}, \(\mathcal{H}_2\) is defined as the total squared strain amplitude:
\begin{align}\label{eq:H2}
  \mathcal{H}_2 
= \sum_{j=1}^{N} \mathcal{A}_j^2 
= \mathcal{A}^2 \sum_{j=1}^{N} j^{-2/3} 
= \mathcal{A}^2 N_s.
\end{align}
Here, \(N_s\) represents the effective number of shells of sources~\cite{Allen2023_Variance}:
\begin{align}\label{eq:Ns}
 N_s = \sum_{n=1}^{N} n^{-2/3} 
 \approx \int_{0}^{N} n^{-2/3} dn \approx 3N^{1/3}.
\end{align}
The fourth-order sum \(\mathcal{H}_4\) is similarly defined as:
\begin{align}\label{eq:H4}
 \mathcal{H}_4 
 = \sum_{j=1}^{N} \mathcal{A}_j^4 
 = \mathcal{A}^4 \sum_{j=1}^{N} j^{-4/3}.
\end{align}
Finally, for the analysis of skewness, we must also define the sixth-order sum, \(\mathcal{H}_6\), which appears in the calculation of the third moment:
\begin{align}\label{eq:H6}
 \mathcal{H}_6 
 = \sum_{j=1}^{N} \mathcal{A}_j^6 
 = \mathcal{A}^6 \sum_{j=1}^{N} j^{-2}.
\end{align}
We note that while \(\mathcal{H}_2\) grows without bound (proportional to \(N^{1/3}\)), both \(\mathcal{H}_4\) and \(\mathcal{H}_6\) converge to constants in the limit of large \(N\). Specifically, \(\mathcal{H}_4\) converges to \(\mathcal{A}^4 \zeta(4/3)\approx3.601\mathcal{A}^4\) and \(\mathcal{H}_6\) converges to \(\mathcal{A}^6 \zeta(2) = \mathcal{A}^6 (\pi^2/6)\approx1.645\mathcal{A}^6\). This convergence of the higher-order sums is a key property of this model~\cite{Allen2023_Variance}.

To define the statistical ensemble for this model, we introduce the following assumptions. We assume that all \(N\) sources radiate at an identical fixed GW angular frequency \(\omega\) and are unpolarized~\cite{Maggiore2000}.
The complex GW strain \(h_j(t)\) for the \(j\)-th source, combining its plus and cross polarization amplitudes, is thus given by:
\begin{align}\label{eq:many strain}
  h_j(t) 
  = h_j^+(t) + ih_j^{\times}(t) 
  = \mathcal{A}_j e^{i(\omega t + \phi_j)}.
\end{align}
The phases \(\phi_j\) are the primary random variables governing the interference. We assume they are statistically independent and uniformly distributed on the interval \([0,2\pi)\). 
The source directions \(\boldsymbol{\Omega}_j\) are also assumed to be independent and uniformly distributed on the sphere.

The ensemble average \(\langle...\rangle\) used in the subsequent moment calculations therefore represents a combined average over these two sets of random variables: the phase \(\langle...\rangle_{\phi}\) and the source directions \(\langle...\rangle_{\boldsymbol{\Omega}}\). The phase average \(\langle...\rangle_{\phi}\) is defined as:
\begin{align}\label{eq:phase average}
 \langle 
  Q
 \rangle_\phi
 =
 \int_0^{2\pi}\frac{d\phi_1}{2\pi}
 ...
 \int_0^{2\pi}\frac{d\phi_N}{2\pi} 
 \,Q(\phi_1,...,\phi_N).
\end{align}
This definition leads to the crucial property \(\langle e^{i(\phi_j - \phi_k)}\rangle_\phi= \delta_{jk}\). This property is the key mechanism that separates the "diagonal" \((j=k)\) terms from the "off-diagonal" \((j\neq k)\) interference terms in the full correlation calculation.

According to the notation in Allen (2023)~\cite{Allen2023_Variance}, we start from the GW-induced redshift \(Z(t)\) for the two pulsars (labeled 1 and 2), which is expressed as a sum over the \(N\) sources:
\begin{align}
\label{eq:many Z1}
  Z_1 (t) 
  &= \sum_j^N
  \left[
   c_j e^{i(\omega t +\phi_j)} +c^*_j e^{-i(\omega t +\phi_j)}
  \right],\\
\label{eq:many Z2}
  Z_2 (t) &
  = \sum_k^N
  \left[
   d_k e^{i(\omega t +\phi_k)} +d^*_k e^{-i(\omega t +\phi_k)}
  \right].
\end{align}
The complex coefficients \(c_j\) and \(d_k\) depend on the \(j\)-th or \(k\)-th source's amplitude \(\mathcal{A}_j\) (or \(\mathcal{A}_k\)), its direction \(\boldsymbol{\Omega}_j\) (or \(\boldsymbol{\Omega}_k\)), the pulsar term parameter \(\chi\), and the respective antenna pattern functions \(F_{1,2}^A\):
\begin{align}
\label{eq:cj}
 c_j 
 &= \frac{1}{2} \mathcal{A}_j
 \left[
  1-\chi e^{i\omega L_1(1+\boldsymbol{p}_1\cdot\boldsymbol{\Omega}_j) }
 \right]
 \left(
  F_1^+(\boldsymbol{\Omega}_j)-iF_1^{\times}(\boldsymbol{\Omega}_j)
 \right),\\
\label{eq:dk}
 d_k 
 &= \frac{1}{2} \mathcal{A}_k
 \left[
  1-\chi e^{i\omega L_2(1+\boldsymbol{p}_2\cdot\boldsymbol{\Omega}_k) }
 \right]
 \left(
  F_2^+(\boldsymbol{\Omega}_k)-iF_2^{\times}(\boldsymbol{\Omega}_k)
 \right).
\end{align}
Before turning to the skewness, we find it instructive to revisit the derivations of the mean and variance, which set the stage for the skewness.

\subsection{\label{sub:many mean}Mean}
The first statistical moment is derived by taking the ensemble average of the product of the time-averaged GW-induced redshifts corresponding to each pulsar, \(\mu = \langle\overline{Z_1Z_2}\rangle\)~\cite{Allen2023_Variance}.
First, the time-averaged product \(\rho = \overline{Z_1Z_2}\) is computed. Under the condition that the angular frequency \(\omega\) is an integer multiple of \(2\pi\) divided by the observation duration \(T\), terms oscillating at \(2\omega t\) vanish upon time averaging~\cite{Allen2023_Variance}. The resulting correlation is then split into "diagonal" \((j=k)\) and "off-diagonal" \((j\neq k)\) terms:
\begin{align} \label{eq:many rho}
 \rho 
 &= \sum_{j,k}
 \left[
  c_{j}d_{k}^* e^{i(\phi_{j}-\phi_{k})}
 +c_{j}^*d_{k} e^{-i(\phi_{j}-\phi_{k})}
 \right]\notag \\   
 &= \rho_\text{diag}
 + \rho_\text{off-diag}
\end{align}
where
\begin{align}
 \rho_\text{diag} 
 &=\sum_i (c_i d_i^* + c_i^* d_i) ,\\
 \rho_\text{off-diag}&=\sum_{i\neq j} 
  \Bigl[
   c_i d_j^* e^{i(\phi_i - \phi_j)} + c_i^* d_j e^{-i(\phi_i - \phi_j)}
  \Bigl].
\end{align}
Next, we apply the ensemble average. The first step is the phase average \(\langle...\rangle_{\phi}\). Due to the property \(\langle e ^ {i(\phi_j- \phi_k)}\rangle_{\phi} = \delta_{jk}\), the phase average of the entire "off-diagonal" \(j\neq k\) sum vanishes. The phase-averaged correlation is therefore determined solely by the "diagonal" terms~\cite{Allen2023_Variance}:
\begin{align}\label{eq:many phase averaged rho}
  \langle
   \rho
  \rangle_\phi 
  =\rho_\text{diag}
  =\sum_j  (c_j d_j^* + c_j^* d_j).
\end{align}
Finally, we compute the ensemble average by taking the average over the source directions \(\boldsymbol{\Omega}_j\). This average, \(\mu = \langle\rho\rangle\equiv \langle\langle\rho\rangle_\phi\rangle_{\boldsymbol{\Omega}}\), requires evaluating \(\langle c_j d_j^*\rangle_{\boldsymbol{\Omega}}\).
As shown in Allen (2023)~\cite{Allen2023_Variance} and earlier works~\cite{HellingsDowns1983}, the rapidly oscillating pulsar terms (those involving \(L_1\) and \(L_2\)) average to zero. Consequently, the product of the pulsar term brackets approximates to unity:
\begin{align}\label{eq:pulsar term corr}
  \left\langle
  \left[
   1-\chi e^{i\omega L_1(1+\boldsymbol{p}_1\cdot\boldsymbol{\Omega}_j) }
  \right]
  \left[
   1-\chi^* e^{-i\omega L_2(1+\boldsymbol{p}_2\cdot\boldsymbol{\Omega}_j) }
  \right] 
  \right\rangle_{\boldsymbol{\Omega}}
  \approx 1.
\end{align}
The imaginary components also vanish upon averaging. The resulting directional average is given by:
\begin{align}\label{eq:integ cjdj*}
 \langle 
  &c_j d_j^* \rangle_{\boldsymbol{\Omega}} \notag\\
 =& \frac{1}{4} \mathcal{A}_j^2 
 \langle 
  (F_1^+ (\boldsymbol{\Omega}_j) + i F_1^\times (\boldsymbol{\Omega}_j)) 
  (F_2^+ (\boldsymbol{\Omega}_j) - i F_2^\times (\boldsymbol{\Omega}_j)) 
 \rangle_{\boldsymbol{\Omega}} \notag\\
 =& \frac{1}{4} \mathcal{A}_j^2 \mu_\text{u}(\gamma).
\end{align}
Substituting this result back into the sum for \(\mu\) and summing over all \(j\) sources, we obtain the mean correlation for the interfering source model~\cite{Allen2023_Variance}:
\begin{align}\label{eq:many mean}
  \mu(\gamma) 
  =\langle\rho\rangle 
  =\frac{1}{2} \mathcal{H}_2 \mu_\text{u}(\gamma).
\end{align}
The mean correlation is directly proportional to the HD curve \(\mu_\text{u}(\gamma)\) defined in Eq.~\eqref{eq:single mean mu} and scales with the total squared strain \(\mathcal{H}_2\) of the \(N\) sources.

\subsection{\label{sub:many var}Variance}
Next, we derive the second moment (variance) of the correlation, \(\sigma^2 = \langle\rho^2\rangle-\mu^2\). The primary task is to compute the ensemble-averaged second moment \(\langle\rho^2\rangle\). We begin by squaring the expression for \(\rho\) from the previous subsection~\cite{Allen2023_Variance}:
\begin{align}\label{eq:many rho^2}
 \rho^2 
 =
 (\rho_\text{diag}+\rho_\text{off-diag})^2.
\end{align}
We first apply the phase average \(\langle...\rangle_\phi\). The cross-term between the diagonal sum and the off-diagonal sum vanishes, as it contains single phase factors like \(\langle e^{i(\phi_j-\phi_k)}\rangle_\phi\), which are zero for \(j\neq k\)~\cite{Allen2023_Variance}. This leaves two terms: the square of the diagonal term (which is independent of phase) and the square of the off-diagonal term:
\begin{align}\label{eq:many phase averaged rho^2}
 \langle
  \rho^2
 \rangle_\phi 
 =& 
 \rho_\text{diag}^2 + \langle\rho_\text{off-diag}^2\rangle_\phi
\end{align}
\begin{widetext}
The phase average of the squared off-diagonal term involves four-phase products:
\begin{align}
 \langle
  \rho^2_\text{off-diag}
 \rangle_\phi 
 =&
\sum_{j\neq k}\sum_{\ell\neq m} 
 \Bigl[ 
  c_j d_k^* c_\ell d_m^* \langle 
   e^{i(\phi_j-\phi_k+\phi_\ell-\phi_m)} 
  \rangle_\phi
  +
  c_j d_k^* c_\ell^* d_m 
  \langle 
   e^{i(\phi_j-\phi_k-\phi_\ell+\phi_m)} 
  \rangle_\phi\notag\\
  &+ 
  c_j^* d_k c_\ell^* d_m 
  \langle 
   e^{-i(\phi_j-\phi_k+\phi_\ell-\phi_m)} 
  \rangle_\phi
  + 
  c_j^* d_k c_\ell d_m^* \langle 
   e^{-i(\phi_j-\phi_k-\phi_\ell+\phi_m)} 
  \rangle_\phi 
 \Bigl].
\end{align}
Since \(j\neq k\) and \(\ell\neq m\), the only non-vanishing contributions arise when the phases cancel out, specifically \(\langle e^{i(\phi_j-\phi_k-\phi_\ell+\phi_m)} \rangle_\phi = \delta_{j\ell}\delta_{km}\). Evaluating these terms simplifies the phase-averaged second moment to~\cite{Allen2023_Variance}:
\begin{align}\label{eq:completed phase averaged rho^2}
  \langle
   \rho^2
  \rangle_\phi 
  &=\sum_j
  \left( 
   c_j d_j^* + c_j^* d_j
  \right)^2
 + 2\sum_{j\neq k}
 \left[
 c_j d_j^* c_k d_k^* +
 c_j^* d_j c_k d_k^* +
 c_j^* d_j c_k^* d_k +
 |c_j|^2|d_k|^2
 \right].
\end{align}
Next, we apply the source direction average \(\langle...\rangle_{\boldsymbol{\Omega}}\). The off-diagonal \(j\neq k\) sum is straightforward as the averages separate into products of first moments, which we found in Sec.~\ref{sec:single}:
\begin{align}\label{eq:two point mean square}
 \langle 
  c_j d_j^* c_k d_k^* \rangle_{\boldsymbol{\Omega}} 
 &= 
 \langle 
  c_j d_j^* \rangle_{\boldsymbol{\Omega}} 
 \langle 
  c_k d_k^* \rangle_{\boldsymbol{\Omega}} 
 = \frac{1}{16}\mathcal{A}_j^2 \mathcal{A}_k^2 \mu_\text{u}^2(\gamma),\\
\label{eq:two point autocorr}
 \langle
  |c_j|^2|d_k|^2
 \rangle_{\boldsymbol{\Omega}} 
 &= 
 \langle
  |c_j|^2
 \rangle_{\boldsymbol{\Omega}} 
 \langle
  |d_k|^2
 \rangle_{\boldsymbol{\Omega}} 
 = \frac{1}{16}\mathcal{A}_j^2 \mathcal{A}_k^2(1+\chi^2)^2 \mu_\text{u}^2(0).
\end{align}
The factor \((1+\chi^2)\) arises from the autocorrelation terms \(\langle|c_j|^2\rangle_{\boldsymbol{\Omega}}\) and \(\langle|d_k|^2\rangle_{\boldsymbol{\Omega}}\), which appear in the \(j\neq k\) (off-diagonal) sum.
When calculating these terms, the magnitude-squared of the pulsar term bracket is given by:
\begin{align}\label{eq:chi square autocorr}
  \left\langle
  \left|
   1-\chi e^{i\omega L_1 (1+\boldsymbol{p}_1\cdot\boldsymbol{\Omega}_j)}
  \right| ^2
  \right\rangle_{\boldsymbol{\Omega}}
  &= 1+ \chi^2 - 2\chi \left\langle \cos(\omega L_1 (1+ \boldsymbol{p}_1\cdot\boldsymbol{\Omega}_j)) \right\rangle_{\boldsymbol{\Omega}} \notag\\
  &\approx 1 +\chi^2.
\end{align}
As argued in Allen (2023)~\cite{Allen2023_Variance}, the rapidly oscillating cosine term averages to zero over source directions (assuming large pulsar distances \(L_1\)), leaving only the \(1+\chi^2\) factor.

The diagonal \((j=k)\) sum requires evaluating the second moments of the single-source correlation. These averages are expressed in terms of the single-source variance functions \(\sigma^2_\text{u}\), \(\sigma^2_\text{p}\), and \(\sigma^2_\text{c}\) defined in Sec.~\ref{sec:single}. The diagonal term is~\cite{Allen2023_Variance}:
\begin{align}\label{eq:integ rho^2 diag}
  \langle
   \left( 
    c_j d_j^* + c_j^* d_j
   \right)^2
  \rangle_{\boldsymbol{\Omega}} 
 &=\langle
  (c_j d_j^*)^2 
 + (c_j^* d_j)^2 
 + 2|c_j|^2|d_j|^2
  \rangle_{\boldsymbol{\Omega}}\notag\\
 &=\frac{1}{8}\mathcal{A}_j^4
  \left(2\mu_\text{u}^2(\gamma) + 
   2\sigma_\text{u}^2(\gamma) + 
   ((1+\chi^2)^2 - 1)\sigma_\text{c}^2(\gamma)
  \right).
\end{align}
We obtain the full second moment \(\langle\rho^2\rangle\) by summing the averaged diagonal terms over \(j\) (proportional to \(\mathcal{H}_4\)) and the averaged off-diagonal terms over \(j\neq k\) (proportional to \(\mathcal{H}_2^2-\mathcal{H}_4\)):
\begin{align}\label{eq:many second moment}
  \langle
   \rho^2
  \rangle 
  &=\frac{1}{8}\mathcal{H}_4
  \left( 
   2\mu_\text{u}^2(\gamma) + 2\sigma_\text{u}^2(\gamma)
 +  \left(
     (1+\chi^2)^2-1
    \right)\sigma_\text{c}^2(\gamma) 
  \right) 
 + \frac{1}{8}(\mathcal{H}_2^2 - \mathcal{H}_4)
 \left(
  3\mu_\text{u}^2(\gamma) + (1+\chi^2)^2 \mu_\text{u}^2(0)
 \right).
\end{align}
Finally, we find the variance \(\sigma^2 = \langle\rho^2\rangle-\mu^2\) by subtracting the square of the mean, \(\mu^2 = \frac{1}{4} \mathcal{H}_2^2 \mu^2_\text{u}(\gamma)\). This cancels several terms, yielding the many-point variance for the interfering case~\cite{Allen2023_Variance}:
\begin{align}\label{eq:many var}
  \sigma^2 
  =\frac{1}{8}\mathcal{H}_4
  \left(
   2\sigma_\text{u}^2(\gamma) 
 +  ((1+\chi^2)^2 - 1)\sigma_\text{c}^2(\gamma)
  \right) 
  + \frac{1}{8}(\mathcal{H}_2^2 - \mathcal{H}_4)
  \left(
   \mu_\text{u}^2(\gamma) 
 +  (1+\chi^2)^2\mu_\text{u}^2(0)
  \right).
\end{align}
\end{widetext}
This is one of the main results of Allen (2023)~\cite{Allen2023_Variance}. In the limit of many sources (\(N\rightarrow\infty\)), the \(\mathcal{H}_2^2\) term dominates; including the pulsar term \((\chi=1)\), the variance simplifies to:
\begin{align}\label{eq:many var approx}
  \sigma^2 \approx \frac{1}{8} \mathcal{H}_2^2 
   \left(
    \mu_\text{u}^2(\gamma) 
 + 4 \mu_\text{u}^2(0)
   \right).
\end{align}

\subsection{\label{sub:many skew}Skewness}
In this section, we derive the third-order moment, or skewness, for the interfering source model. This is the primary result of this section. The skewness is defined by the normalized third cumulant \(\mathcal{S} = \kappa_3/\sigma^3\), which characterizes the non-Gaussianity of the background~\cite{Bartolo2018}. Our central task is to compute the third cumulant, \(\kappa_3\), defined as:
\begin{align}\label{eq:kappa}
  \kappa_3 = 
  \langle
   \rho^3
  \rangle 
  -3\mu\sigma^2 -\mu^3.
\end{align}
This requires the full ensemble average of the third moment, \(\langle\rho^3\rangle\). We begin by cubing the full correlation:
\begin{align}\label{eq:rho^3}
    \rho^3 = 
    \bigl(
    \rho_\text{diag} + \rho_\text{off-diag}
    \bigl)^3,
\end{align}
We first apply the phase average \(\langle...\rangle_\phi\) following the methodology of Allen (2023)~\cite{Allen2023_Variance}. Upon expanding Eq.~\eqref{eq:rho^3}, the term \(3\langle \rho_\text{diag}^2\rho_\text{off-diag}\rangle_\phi\) vanishes because it contains odd powers of the off-diagonal phase factors. Analogous to the cross-term in the variance calculations, its mean is zero. This leaves three non-vanishing terms:
\begin{align}\label{eq:phase averaged rho^3}
  \langle
   \rho^3
  \rangle_\phi
  = 
  \langle
   \rho_\text{diag}^3
  \rangle_\phi 
+3\langle
   \rho_\text{diag}\rho_\text{off-diag}^2
  \rangle_\phi
+ \langle
   \rho_\text{off-diag}^3
  \rangle_\phi.
\end{align}
We now analyze the full ensemble average (phase and direction) of each of these terms. The exact analytical expression for \(\langle\rho^3\rangle\) is a complex sum of terms involving different index combinations of the amplitude sums, resulting in contributions proportional to \(\mathcal{H}_6\), \(\mathcal{H}_4\mathcal{H}_2\), and \(\mathcal{H}_2^3\). 
\begin{itemize}
  \item \textbf{Non-dominant terms:} The \(\mathcal{H}_6\) and \(\mathcal{H}_4\mathcal{H}_2\) terms arise from index contractions where two or more source indices are the same (e.g., \(i=j=k\) or \(i=j\neq k\)). These terms are significant only when \(N\) is small.
  \item \textbf{Dominant terms:} The \(\mathcal{H}_2^3\) terms arise when all relevant source indices are distinct (e.g., \(i\neq j\neq k\)).
\end{itemize}
Since we are interested in the limit of many sources (\(N \rightarrow \infty\)), where \(\mathcal{H}_2\propto N^{1/3}\) diverges while \(\mathcal{H}_4\) and \(\mathcal{H}_6\) converge to constants~\cite{Allen2023_Variance}, we will focus on the dominant terms proportional to \(\mathcal{H}_2^3\) and omit the detailed calculation of the non-dominant terms.

\begin{enumerate}
  \item \textbf{The \(\langle\rho_\text{diag}^3\rangle\) Term}
  
 This term, \(\left\langle \left(\sum_j (c_j d_j^* + c_j^* d_j)\right)^3\right\rangle\), is independent of phase. The expansion contains terms proportional to \(\mathcal{H}_6\), \(\mathcal{H}_4 \mathcal{H}_2\), and \(\mathcal{H}_2^3\). In the large-\(N\) limit, the \(i \neq j\neq k\) term dominates. Performing the direction average on this dominant term yields:
  \begin{align}\label{eq:rho^3-diag}
   \langle\rho_\text{diag}^3\rangle 
    &\approx 
    \left\langle
      \sum_{i\neq j\neq k}
      (c_i d_i^* + c_i^* d_i)
      (c_j d_j^* + c_j^* d_j)
      (c_k d_k^* + c_k^* d_k)
    \right\rangle_{\boldsymbol{\Omega}}\notag\\
    &=\frac{1}{8}\mathcal{H}_2^3\mu_\text{u}^3(\gamma).
  \end{align}

  \item \textbf{The \(\langle3\rho_\text{diag}\rho_\text{off-diag}^2\rangle\) Term}
  
  Due to the complexity of this calculation, the detailed steps are presented in Appendix~\ref{app:many}. The dominant term in the result is given by:
  \begin{align}
        \left\langle
   3\rho_\text{diag}\rho_\text{off-diag}^2
  \right\rangle
 &\approx\frac{3}{16}
  \mathcal{H}_2^3
  \Bigl[
   \mu_\text{u}^3(\gamma) 
+  \mu_\text{u}(\gamma)(1+\chi^2)^2\mu_\text{u}^2(0)
  \Bigl].
  \end{align}

  \item \textbf{The \(\langle\rho_\text{off-diag}^3\rangle\) Term}
  
  The detailed calculation for this term is also provided in Appendix~\ref{app:many}. The dominant term is given by:
  \begin{align}
       \langle
   \rho_\text{off-diag}^3
  \rangle
 &\approx
 \frac{1}{16} \mathcal{H}_2^3
 \Bigl[
  \mu_\text{u}^3(\gamma) 
+3\mu_\text{u}(\gamma)(1+\chi^2)^2\mu_\text{u}^2(0)
 \Bigl].
  \end{align}
\end{enumerate}

\begin{figure}[t]
  \centering
  \includegraphics[width=0.45\textwidth]{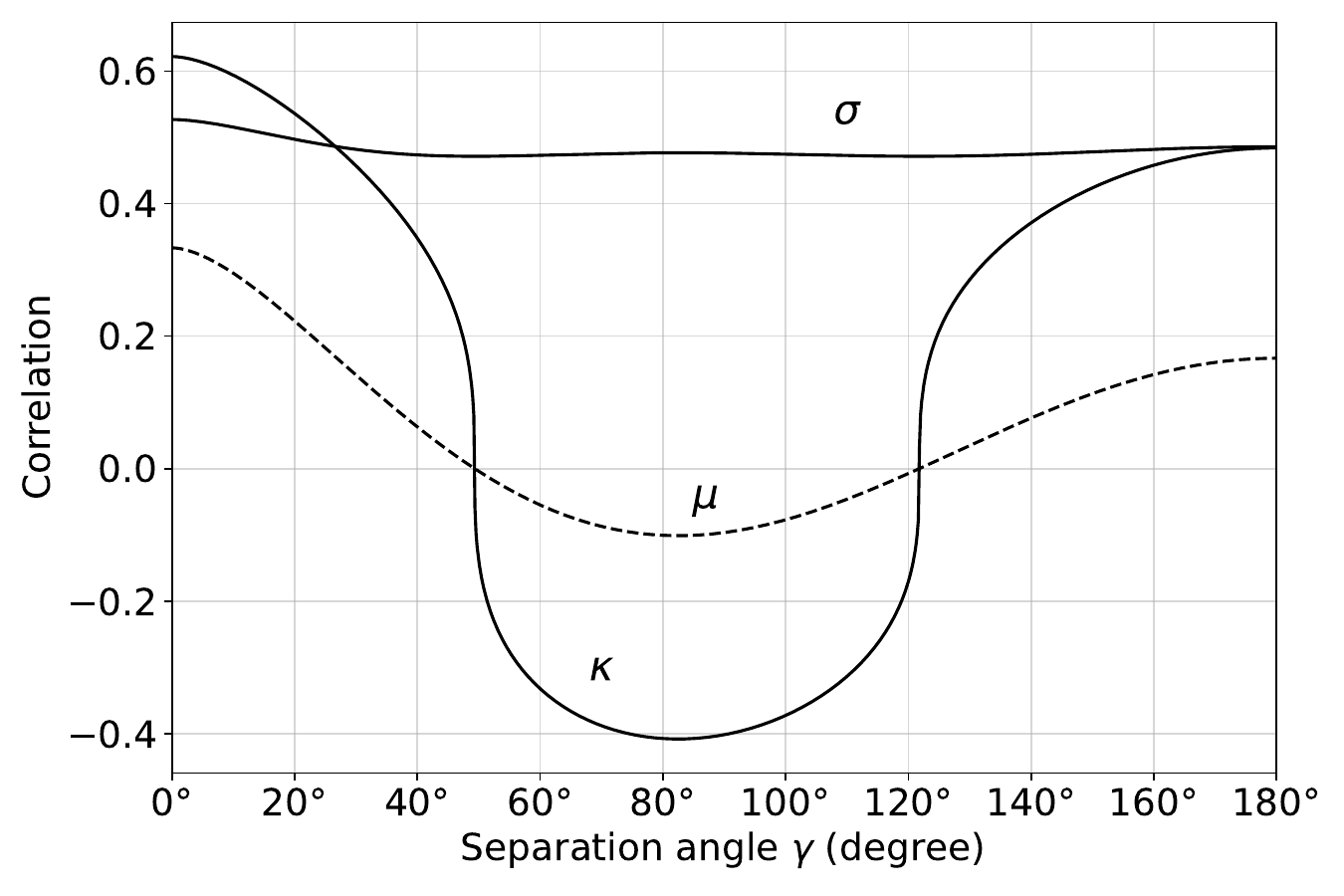}
  \caption{Statistical quantities for a large number of sources where \(\mathcal{H}_2^2 \gg \mathcal{H}_4\). We assume the interfering source model and set the pulsar term \(\chi=1\). The dashed line represents the many-point mean \(\mu\) (equivalent to the single-source mean without normalization), the solid lines show the many-point standard deviation \(\sigma\) and the cubic root of the skewness numerator \(\kappa\).}
  \label{fig:many_mu_sigma_kappa}
\end{figure}

Finally, we combine the dominant contributions derived from the three terms: \(\langle\rho_\text{diag}^3\rangle\), \(3\langle\rho_\text{diag}\rho_\text{off-diag}^2\rangle\), and \(\langle\rho_\text{off-diag}^3\rangle\). Summing the \(\mathcal{H}^3_2\) coefficients from each term yields the total ensemble-averaged third moment in the large-\(N\) limit:
\begin{align} \label{eq:dom rho^3}
 \langle\rho^{3}\rangle
 &\approx \frac{3}{8}\mathcal{H}_2^3
 \Bigl[
  \mu_\text{u}^3(\gamma)  
 +\mu_\text{u}(\gamma)(1+\chi^2)^2\mu_\text{u}^2(0) 
 \Bigl].
\end{align}
To obtain the physical skewness, we must calculate the third central moment \(\kappa_3\), which is defined in Eq.~\eqref{eq:kappa}.
We substitute the large-\(N\) approximations for the mean \(\mu\) Eq.~\eqref{eq:many mean} and variance \(\sigma^2\) Eq.~\eqref{eq:many var approx} into this definition:
\begin{itemize}
 \item Mean cubed: 
 \begin{align}
     \mu^3 = \frac{1}{8}\mathcal{H}_2^3\mu_\text{u}^3(\gamma).
 \end{align}
 \item Cross term: 
 \begin{align}
     3\mu\sigma^2 \approx \frac{3}{16}\mathcal{H}_2^3\biggl[\mu_\text{u}^3(\gamma)+ \mu_\text{u}(\gamma) (1+\chi^2)^2\mu_\text{u}^2(0) \biggl].
 \end{align}
\end{itemize}
Subtracting these from the dominant term of \(\langle\rho^3\rangle\) yields the final expression for the cumulant.
If we properly include the pulsar term by setting (\(\chi=1\)), the dominant non-Gaussian component of the correlation is:
\begin{align}
 \kappa_3 \approx 
 \frac{1}{16}\mathcal{H}_2^3
 \Bigl[
  \mu_\text{u}^3(\gamma)  
  +12\mu_\text{u}(\gamma)\mu_\text{u}^2(0) 
 \Bigl].
\end{align}This result indicates that a non-zero skewness survives in the large-\(N\) limit of the interfering source model. 
The sign and shape of the skewness are determined by the cubic HD curve \(\mu^3_\text{u}(\gamma)\) and the cross-term scaled by the monopole moment \(\mu_\text{u}(0)\).
For visualization, we define the cubic root of the many-point skewness numerator as follows, similar to Eq.~\eqref{eq:root kappa}:
\begin{align}
    \kappa=\sqrt[3]{\kappa_3}.
\end{align}
We plot the many point mean, standard deviation, and the cubic root of the skewness numerator in Fig.~\ref{fig:many_mu_sigma_kappa}.

\begin{figure}[t]
  \centering
  \includegraphics[width=0.45\textwidth]{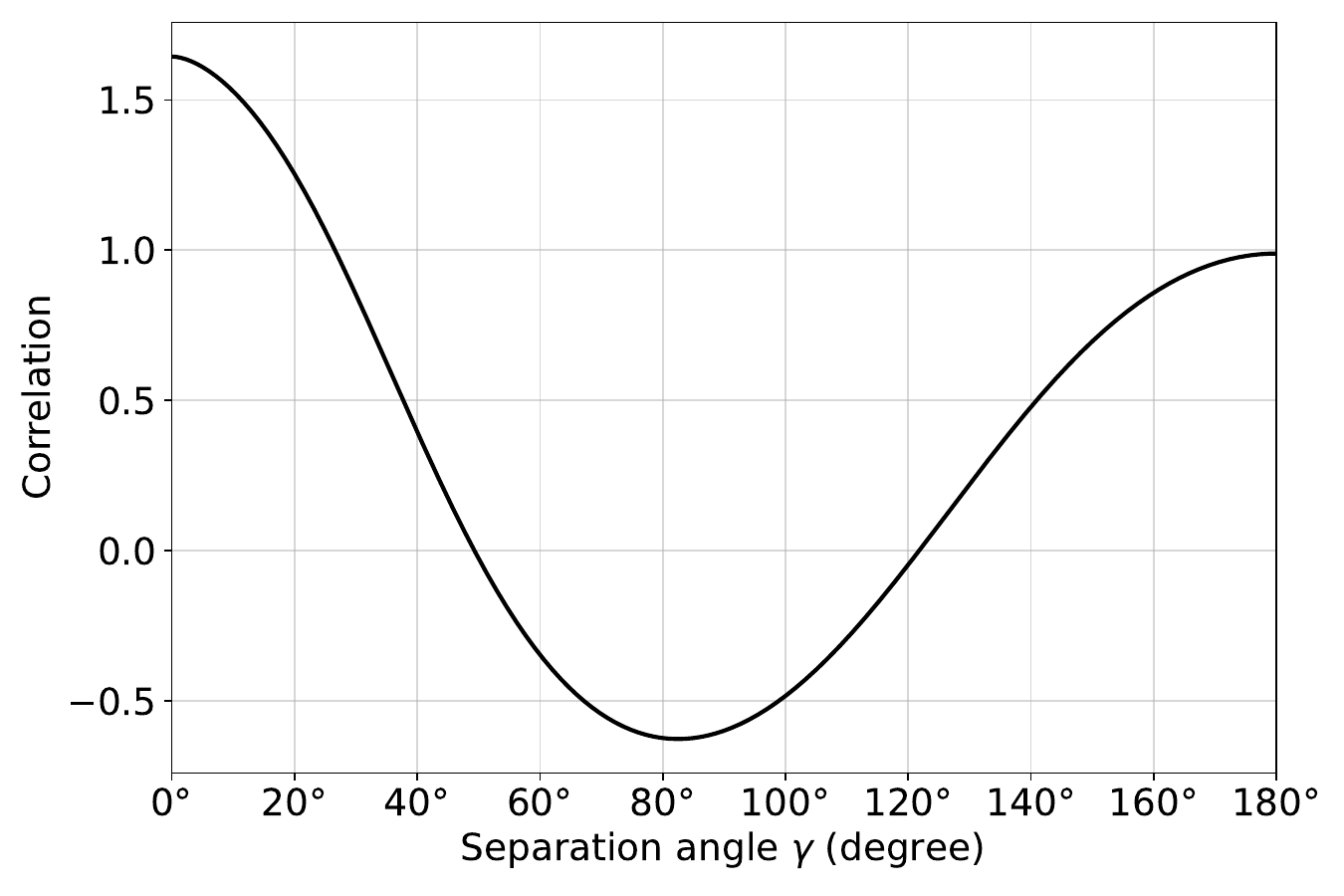}
  \caption{The many-point skewness \(\mathcal{S}\) as a function of the angular separation \(\gamma\).}
  \label{fig:many_norm_skew}
\end{figure}

Finally, we obtain the many point skewness \(\mathcal{S}(\gamma) = \kappa_3/\sigma^3\) by combining this result with the variance derived in Eq.~\eqref{eq:many var approx}. Notably, the source amplitude factors \(\mathcal{H}_2\) cancel out, leaving a scale-independent shape function:
\begin{align}\label{eq:many_skew_final}
 \mathcal{S}(\gamma)
 &\approx \frac{\frac{1}{16}\mathcal{H}_2^3 \left[ \mu_\text{u}^3(\gamma) + 12\mu_\text{u}(\gamma)\mu_\text{u}^2(0) \right]}
 {\left( \frac{1}{8}\mathcal{H}_2^2 \left[ \mu_\text{u}^2(\gamma) + 4\mu_\text{u}^2(0) \right] \right)^{3/2}} \notag\\
 &= \sqrt{2} \frac{\mu_\text{u}^3(\gamma) + 12\mu_\text{u}(\gamma)\mu_\text{u}^2(0)}{\left( \mu_\text{u}^2(\gamma) + 4\mu_\text{u}^2(0) \right)^{3/2}}.
\end{align}
This expression quantifies the intrinsic non-Gaussianity of the interfering background, independent of the total GW intensity. As shown in Fig.~\ref{fig:many_norm_skew}, the many-point skewness $S(\gamma)$ is an $\mathcal{O}(1)$ function that is positive at small and large separations and negative at intermediate angles, with zero-crossings aligned with those of the HD mean. This follows directly from Eq.~\ref{eq:many_skew_final}, which factorizes as $\mathcal{S}(\gamma)\propto \mu_u(\gamma)$ (the remaining $\gamma$-dependent prefactor is non-negative), so the sign of $\mathcal{S}(\gamma)$ tracks the sign of $\mu_u(\gamma)$.

\section{\label{sec:cosmic}Cosmic Moments}
In the previous section, we analyzed the statistical properties of the correlation \(\rho\) for individual pulsar pairs. The variance and skewness derived there represent the total fluctuations, which arise from two distinct physical origins: "pulsar variance" and "cosmic variance"~\cite{Roebber2016}.
Pulsar variance originates from the specific geometric configuration of pulsar pairs on the sky relative to the GW sources. In contrast, cosmic variance arises from the interference between GW sources radiating at the same frequency, which creates specific interference patterns on the sky that deviate from the isotropic Helling-Downs prediction~\cite{TaylorGair2013, Mingarelli2013,Cornish:2014,Agazie_2023_Anisotropy,Agazie_2024_Polarization, 2022JCAP...11..046B}.

To isolate the intrinsic statistical properties of the GW background itself, specifically the cosmic skewness, we must remove the contribution of pulsar variance. Following the methodology of Allen (2023)~\cite{Allen2023_Variance}, we achieve this by introducing the \textit{pulsar-averaged correlation}, \(\Gamma(\gamma)\). \(\Gamma(\gamma)\) is defined by averaging the correlation \(\rho\) over all possible pairs of pulsars separated by a fixed angle \(\gamma\)~\cite{AllenRomano2023}. This averaging procedure eliminates the pulsar effect. In other words, it averages out the specific geometric orientation of the pulsar pairs, leaving a quantity that depends only on the source interference pattern.

For the interfering source model discussed in Sec.~\ref{sec:many}, the pulsar-averaged correlation \(\Gamma(\gamma)\) is derived by averaging the redshift correlation Eq.~\eqref{eq:many rho} over pulsar positions. As detailed in the calculation of cosmic variance in Allen (2023)~\cite{Allen2023_Variance}, this yields:
\begin{align}\label{eq:Gamma}
 \Gamma(\gamma) 
 =& 
 \frac{1}{4}\sum_{j, k}\mathcal{A}_{j}\mathcal{A}_{k}
 \left(
  e^{i(\phi_{j}-\phi_{k})}+e^{-i(\phi_{j}-\phi_{k})}
 \right)\mu(\gamma,\beta_{jk}) \notag\\
 =& 
 \frac{1}{2}\mathcal{H}_2\mu_\text{u}(\gamma) \notag\\
 &+ 
 \frac{1}{4}\sum_{j\ne k}\mathcal{A}_{j}\mathcal{A}_{k}
 \left(
  e^{i(\phi_{j}-\phi_{k})}+e^{-i(\phi_{j}-\phi_{k})}
 \right)\mu(\gamma,\beta_{jk}).
\end{align}
Here, \(\mu(\gamma, \beta_{jk})\) is the "two-point function" derived in Allen (2023)~\cite{Allen2023_Variance} and Allen \& Romano (2023)~\cite{AllenRomano2023}, which describes the correlation between pulsars separated by \(\gamma\) due to two GW sources separated by an angle \(\beta_{jk}\). The term \(\beta_{jk}\) represents the angle between the directions of source \(j\) and source \(k\). The first term in Eq.~\eqref{eq:Gamma} corresponds to the standard HD correlation, while the second term represents the fluctuations due to source interference. In the following, again, we start from mean and variance, and then proceed to skewness.

\subsection{\label{sub:cosmic mean}Mean in the Pulsar-Averaged Correlation}
The first moment of the pulsar-averaged correlation is obtained by taking the ensemble average over the random phases \(\phi\). Since the sources have independent random phases, the phase average of the interference term where \(j\neq k\) vanishes because \(\langle e^{i(\phi_j-\phi_k)}\rangle_\phi =0\). Thus, the mean of the pulsar-averaged correlation is determined solely by the first term of Eq.~\eqref{eq:Gamma}~\cite{Allen2023_Variance}:
\begin{align}\label{eq:cosmic mean}
 \mu
 = 
 \langle 
  \Gamma(\gamma) 
 \rangle 
 = 
 \frac{1}{2}\mathcal{H}_2 \mu_\text{u}(\gamma).
\end{align}
This result confirms that, on average, the correlation follows the standard HD curve \(\mu_\text{u}(\gamma)\), scaled by the total intensity \(\mathcal{H}_2\)~\cite{HellingsDowns1983}. This matches the mean derived in the previous section, as averaging over pulsars does not change the ensemble mean.

\subsection{\label{sub:cosmic var}Cosmic Variance}
The cosmic variance, \(\sigma_\text{cosmic}^2\), quantifies the deviation of the pulsar-averaged correlation \(\Gamma(\gamma)\) from the mean due to the specific interference pattern of the sources~\cite{Roebber2016}. It is defined as:
\begin{align}
 \sigma_\text{cosmic}^{2}
 =\langle\Gamma^{2}(\gamma)\rangle-\mu^{2}.
\end{align}
To compute the second moment \(\langle \Gamma^2(\gamma) \rangle\), we square Eq.~\eqref{eq:Gamma}. Upon phase averaging, the cross-terms between the diagonal part and the off-diagonal part vanish. The square of the interference sum yields terms involving \(\delta_{jm}\delta_{k\ell}\) and \(\delta_{j\ell}\delta_{km}\). Summing over the distinct indices and averaging over the source positions---which corresponds to averaging the two-point function over \(\beta_{jk}\)---we obtain~\cite{AllenRomano2023}:
\begin{widetext}
\begin{align}
 \langle
  \Gamma^{2}(\gamma)
 \rangle
 &=
 \frac{1}{4}\mathcal{H}_{2}^{2}
 \mu_\text{u}^{2}(\gamma)+
 \frac{1}{8}
 \sum_{j\ne k}\sum_{\ell\ne m}
 \mathcal{A}_{j}\mathcal{A}_{k}\mathcal{A}_{\ell}\mathcal{A}_{m}
 (\delta_{jm}\delta_{k\ell}+\delta_{j\ell}\delta_{km})
 \langle
  \mu(\gamma,\beta_{jk})\mu(\gamma,\beta_{\ell m})
 \rangle_{\boldsymbol{\Omega}}
 \notag\\
 &=
 \frac{1}{4}\mathcal{H}_{2}^{2}
 \mu_\text{u}^{2}(\gamma)+
 \frac{1}{4}
 \sum_{j\ne k}\mathcal{A}_{j}^{2}\mathcal{A}_{k}^{2}
 \langle
  \mu^2(\gamma,\beta_{jk})
 \rangle_{\boldsymbol{\Omega}}\notag\\
 &=
 \frac{1}{4}\mathcal{H}_{2}^{2}
 \mu_\text{u}^{2}(\gamma)+
 \frac{1}{4}
 \sum_{j\ne k}\mathcal{A}_{j}^{2}\mathcal{A}_{k}^{2}
 \widetilde{\mu^2}(\gamma),
\end{align}  
\end{widetext}
where \(\widetilde{\mu^2}(\gamma)\) is the average of the squared two-point function \(\mu^2(\gamma, \beta)\) over the separation angle \(\beta\). The detailed computation of this average is shown in Appendix G of Allen (2023)~\cite{Allen2023_Variance}.
Subtracting the square of the mean from Eq.~\eqref{eq:cosmic mean} yields the cosmic variance:
\begin{align}\label{eq:cosmic variance}
 \sigma^{2}_\text{cosmic}
 &=\langle\Gamma^{2}(\gamma)\rangle-\mu^{2}\nonumber\\
 &=
 \frac{1}{4}
 \sum_{j\ne k}\mathcal{A}_{j}^{2}\mathcal{A}_{k}^{2}
 \widetilde{\mu^2}(\gamma)
 =\frac{1}{4}
 (\mathcal{H}_{2}^{2}-\mathcal{H}_{4})
 \widetilde{\mu^{2}}(\gamma).
\end{align}
This result shows that the cosmic variance scales with \(\mathcal{H}_2^2 - \mathcal{H}_4\), which is dominated by \(\mathcal{H}_2^2\) in the large-\(N\) limit~\cite{Allen2023_Variance}.

\subsection{\label{sub:cosmic skew}Cosmic Skewness}
Finally, we calculate the third central moment, which characterizes the non-Gaussian asymmetry of the pulsar-averaged correlation~\cite{Bartolo2018}. It is defined as:
\begin{align}\label{eq-def-cosmic-kappa}
 \kappa_{3,\text{cosmic}}
 &=
 \langle
  (\Gamma(\gamma)-\mu)^{3}
 \rangle\nonumber\\
 &=
 \langle
  \Gamma^{3}(\gamma)
 \rangle-
 3\mu\sigma_\text{cosmic}^{2}-
 \mu^{3}.
\end{align}
Calculation of the third moment \(\langle \Gamma^3(\gamma) \rangle\) requires cubing Eq.~\eqref{eq:Gamma}. The phase averaging selects specific combinations of indices \(i, j, k\) in the triple sum that form closed phase loops. The result of phase averaging the third moment is:
\begin{align}
  \langle 
   (\Gamma(\gamma))^3
  \rangle_\phi
  =&
  \frac{1}{8}\mathcal{H}_2^3\mu_\text{u}^3(\gamma)\notag\\
  &+
  \frac{3}{8}\sum_i\sum_{j\neq k}
  \mathcal{A}_i^2\mathcal{A}_j^2\mathcal{A}_k^2 
  \mu_\text{u}(\gamma) \mu^2(\gamma,\beta_{jk})\notag\\
  &+
  \frac{1}{4}\sum_{i\neq j\neq k}
  \mathcal{A}_i^2\mathcal{A}_j^2\mathcal{A}_k^2 
  \mu(\gamma,\beta_{ij})
  \mu(\gamma,\beta_{ki})
  \mu(\gamma,\beta_{jk}),
\end{align}
where we employ the same combination of Kronecker delta terms as in Eq.~\eqref{delta_combination}, and utilize the symmetry relationship \(\mu(\gamma,\beta_{jk})=\mu(\gamma,\beta_{kj})\) established in Allen \& Romano (2023)~\cite{AllenRomano2023}.

Using the summation identities:
\begin{align}
    \sum_{i}\sum_{j\neq k} \mathcal{A}^2_i\mathcal{A}^2_j\mathcal{A}^2_k 
    &= \mathcal{H}_2^3-\mathcal{H}_4\mathcal{H}_2,
\end{align}
and
\begin{align}
    \sum_{i \neq j\neq k} \mathcal{A}^2_i\mathcal{A}^2_j\mathcal{A}^2_k 
    &= \mathcal{H}_2^3-3\mathcal{H}_4\mathcal{H}_2+ 2\mathcal{H}_6,
\end{align}
the full ensemble average is given by:
\begin{align}
 \langle
  \Gamma(\gamma)^3
 \rangle
 &=
 \frac{1}{8}\mathcal{H}_{2}^{3}
 \mu_{u}^{3}(\gamma)+
 \frac{3}{8}
 (\mathcal{H}_{2}^{3}-\mathcal{H}_{4}\mathcal{H}_{2})
 \mu_{u}(\gamma)
 \widetilde{\mu^2}(\gamma)\nonumber\\
 &+
 \frac{1}{4}(\mathcal{H}_{2}^{3}-
 3\mathcal{H}_{4}\mathcal{H}_{2}+
 2\mathcal{H}_{6})\widehat{\mu^3}(\gamma).
\end{align}
Here, we have introduced the \textit{three-point-average function} \(\widehat{\mu^3}(\gamma)\), which represents the correlation averaged over three independent GW sources. It is defined by averaging over the relative configurations of three sources, as illustrated in Fig.~\ref{fig:cosmic_set}:
\begin{align}\label{eq:three-point-average function}
 \widehat{\mu^3}(\gamma)
 =
 \langle
  \mu(\gamma,\beta_{ij})
  \mu(\gamma,\beta_{ki})
  \mu(\gamma,\beta_{jk})
 \rangle_{\boldsymbol{\Omega}}.
\end{align}

\begin{figure}[t]
  \centering
  \includegraphics[width=0.45\textwidth]{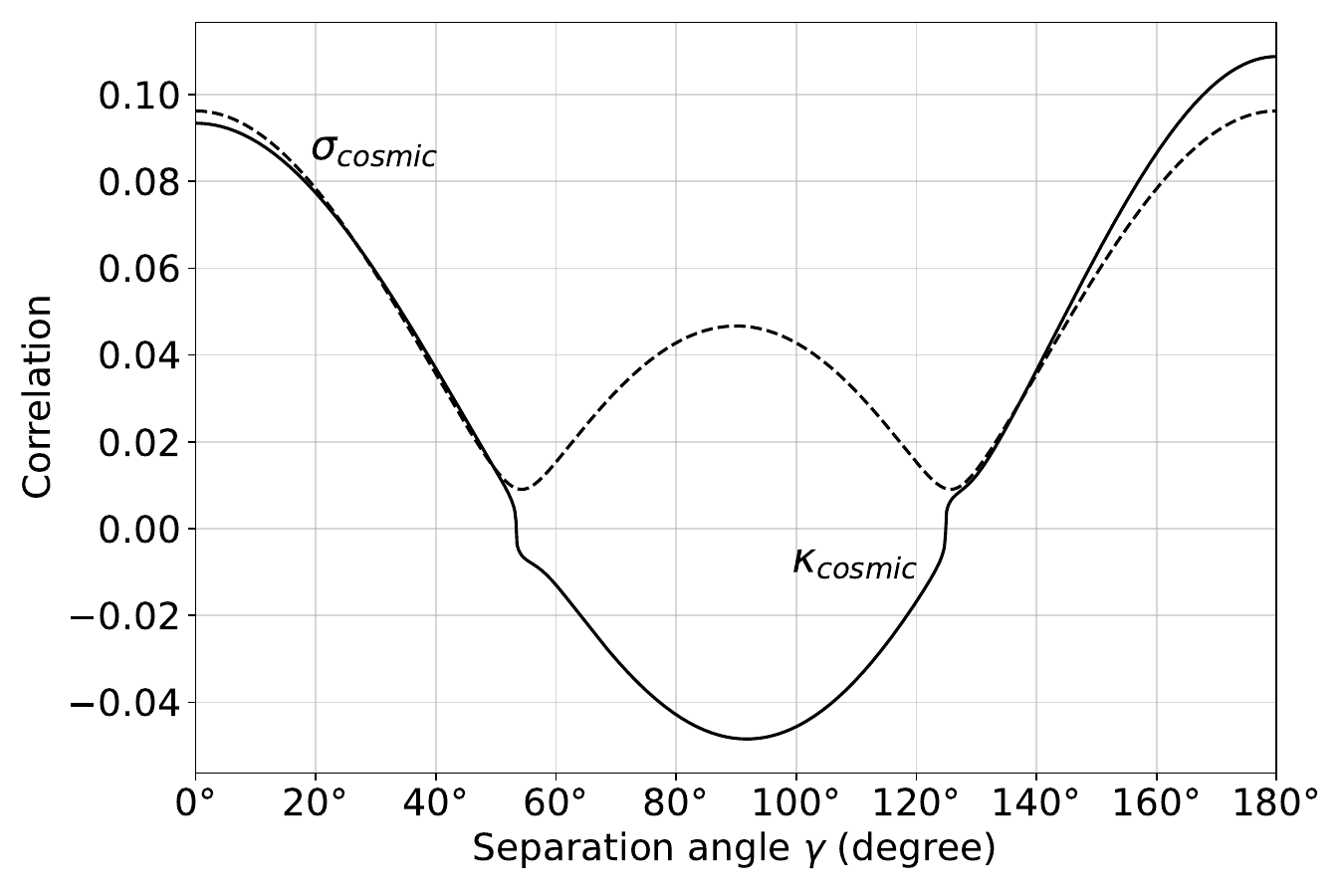}
  \caption{Comparison of cosmic fluctuations. The solid line represents the cubic root of the cosmic skewness numerator \(\kappa_\text{cosmic}\), and the dashed line shows the square root of the cosmic variance \(\sigma_\text{cosmic}\).}
  \label{fig:cosmic_sigma_kappa}
\end{figure}

Substituting the expressions for the mean, variance, and third moment back into Eq.~\eqref{eq-def-cosmic-kappa}, the lower-order terms proportional to \(\mu_\text{u}^3\) and \(\mu_\text{u}\tilde{\mu^2}\) cancel out. This cancellation isolates the purely non-Gaussian contribution from the three-point interaction:
\begin{align}\label{eq:kappa 3 cosmic}
 \kappa_{3,\text{cosmic}}
 =
 \frac{1}{4}
 (\mathcal{H}_{2}^{3}-
 3\mathcal{H}_{4}\mathcal{H}_{2}+
 2\mathcal{H}_{6})
 \widehat{\mu^3}(\gamma).
\end{align}
For visualization purposes, we also define the cubic root of the cosmic skewness numerator:
\begin{align}
    \kappa_\text{cosmic} =\sqrt[3]{\kappa_{3,\text{cosmic}}}.
\end{align}
We plot the square root of the cosmic variance and the cubic root of the cosmic skewness numerator in Fig.~\ref{fig:cosmic_sigma_kappa}.

\begin{figure}[t]
  \centering
  \includegraphics[width=0.45\textwidth]{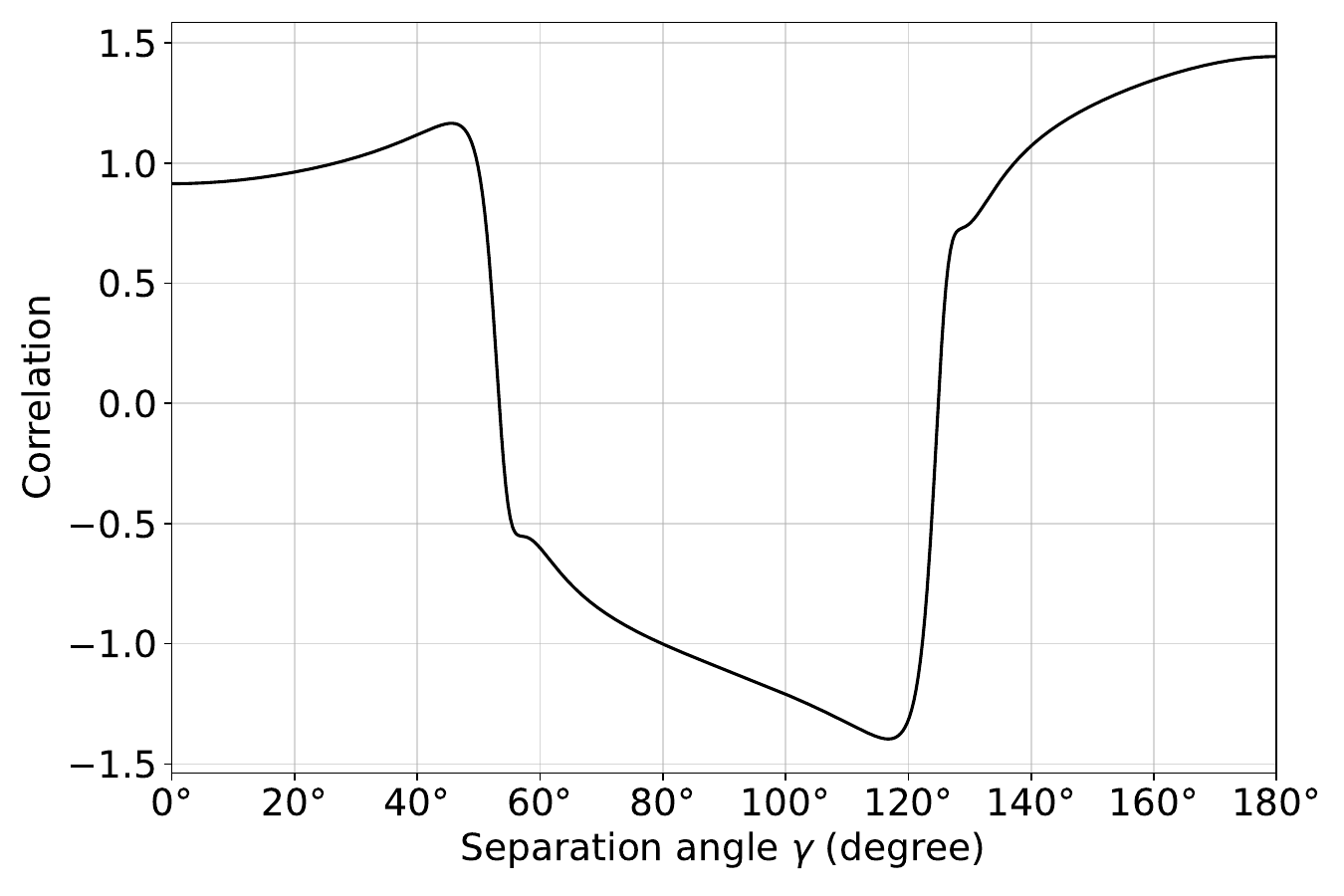}
  \caption{The cosmic skewness \(\mathcal{S}_\text{cosmic}\) as a function of angular separation \(\gamma\).}
  \label{fig:cosmic_norm_skew}
\end{figure}

Finally, dividing the cosmic skewness numerator by the cosmic variance raised to the power of \(3/2\) yields the cosmic skewness:
\begin{align}\label{eq:cosmic skewness}
    \mathcal{S}_\text{cosmic}
    &= \frac{\kappa_{3,\text{cosmic}}}{(\sigma^2_\text{cosmic})^{3/2}} \nonumber \\
    &= 2 \frac{\mathcal{H}_{2}^{3}-
    3\mathcal{H}_{4}\mathcal{H}_{2}+
    2\mathcal{H}_{6}}{(\mathcal{H}_{2}^{2}-\mathcal{H}_{4})^{3/2}}
    \frac{\widehat{\mu^3}(\gamma)}{(\widetilde{\mu^{2}}(\gamma))^{3/2}}.
\end{align}
In the large-$N$ limit, this reduces to:
\begin{align}
    \mathcal{S}_\text{cosmic}
    \approx 2 
    \frac{\widehat{\mu^3}(\gamma)}{(\widetilde{\mu^{2}}(\gamma))^{3/2}}.
\end{align}
The behavior of the cosmic skewness is illustrated in Fig.~\ref{fig:cosmic_norm_skew}.

This result demonstrates that the cosmic skewness is non-zero for \(N_\text{source}\geq 3\) and depends explicitly on the three-point function \(\widehat{\mu^3}(\gamma)\). Unlike the variance, which depends on pairwise source separations, the skewness probes the higher-order statistical structure of the GW background via triplets of interfering sources. The detailed calculation of \(\widehat{\mu^3}(\gamma)\) involves averaging over the geometric configurations of three sources and is discussed in Appendix~\ref{app:cosmic}.

\section{\label{sec:Discussion and Summary}Discussion and Summary}

\begin{figure}[t]
  \centering
  \includegraphics[width=0.5\textwidth]{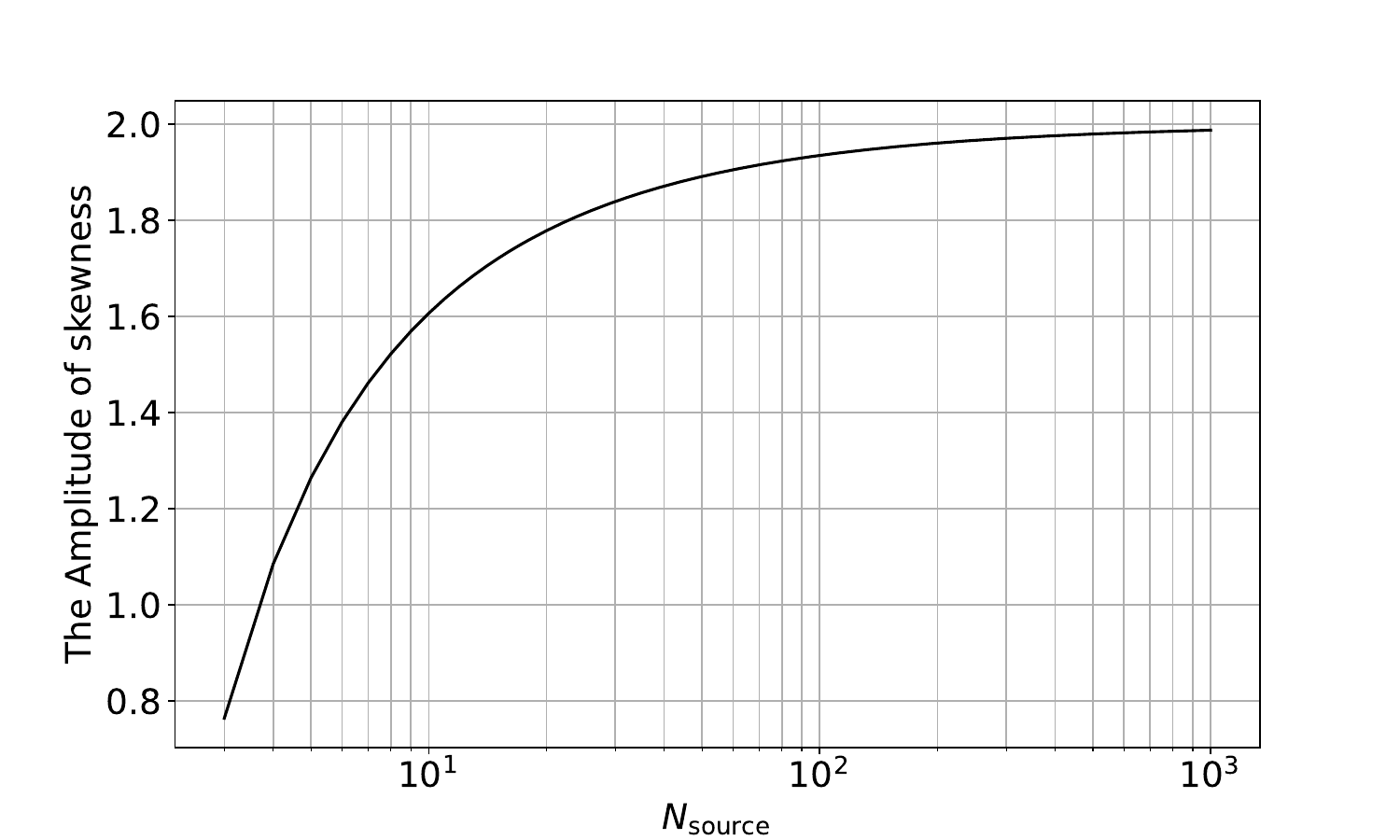}
  \caption{The amplitude of cosmic skewness, defined as $~2 (\mathcal{H}_{2}^{3}-
    3\mathcal{H}_{4}\mathcal{H}_{2}+
    2\mathcal{H}_{6})/{(\mathcal{H}_{2}^{2}-\mathcal{H}_{4})^{3/2}}$ as a function of $N_\text{source}$. This corresponds to the coefficient in Eq.~\eqref{eq:cosmic skewness}.}
  \label{fig:cosmic_skew_amplitude}
\end{figure}

In this paper, we have presented a comprehensive statistical analysis of the HD correlation, focusing specifically on the third-order moments that characterize the fluctuations and non-Gaussian features of the SGWB. Building upon the theoretical framework of variance calculation established by Allen (2023)~\cite{Allen2023_Variance}, we extended the analysis to the third central moment, or skewness, to probe the asymmetry of the correlation distribution.

We first derived the analytical expressions for the skewness induced by a single GW source and subsequently extended the analysis to the "interfering source model" involving a superposition of many discrete sources.
A central finding of this study is that the skewness does not vanish even in the limit of a large number of sources ($N \to \infty$). We demonstrated that in this large-$N$ limit, the third cumulant scales with the cube of the total intensity $\mathcal{H}_2^3$. Consequently, the normalized skewness converges to a non-zero function that becomes independent of the intensity scale. This indicates that the SGWB formed by interfering discrete sources possesses an intrinsic non-Gaussian nature that persists even in the high-source-count regime.

Furthermore, we successfully isolated the "Cosmic Skewness" by removing the pulsar variance contribution to capture the fluctuations inherent to the background interference itself. To describe this quantity, we introduced the "three-point-average function" $\widehat{\mu^3}(\gamma)$, a new statistical function defined by averaging the correlation over the configurations of three independent GW sources.

According to the results of this paper, both the total skewness and the cosmic skewness have the same sign as the mean (the HD correlation). The total skewness is proportional to the HD correlation itself, while the sign of the cosmic skewness is determined by $\widehat{\mu^3}(\gamma)$, which also has almost the same sign as the HD correlation. In this case, the characteristic quantities of the probability distribution are alined as,
\begin{equation}
|\mathrm{mode}| < |\mathrm{median}| < |\mathrm{mean}|,
\end{equation}
for both positive and negative sides of correlation. Moreover, the tail of the probability distribution extends in the direction of increasing absolute correlation. Consequently, most correlation values obtained in observations are closer to zero than the mean, whereas rare occurrences of large correlations pull the mean, manifesting as an asymmetry of the distribution.

In a standard pipeline of gravitational-wave analysis, timing residuals are modeled as a multivariate Gaussian process and inference is performed using a Gaussian likelihood. In this framework, the spatial correlations of the common red process, including a SGWB, are encoded by the overlap reduction function (ORF), $\Gamma_{ab}$, entering the Fourier-domain covariance blocks. Multiple ORF hypotheses are considered in practice: not only the HD correlation expected for a SGWB, but also a monopolar ORF associated with clock errors and a dipolar ORF associated with Solar System ephemeris errors, and they are compared via Bayes factors. Related in spirit, Bernardo, and Ng \cite{BernardoLiuNg2023} investigated the impact of cosmic variance as a theoretical uncertainty in PTA analyses at the level of noise-marginalized spatial cross-correlation measurements, i.e., using a correlation-level likelihood rather than the full residual-level likelihood.

Motivated by the results of this work, one may naturally propose a hierarchical Bayesian model in which the gravitational-wave ORF is not fixed to the HD form but treated as a random variable. Concretely, conditional on a particular realization of the ORF (i.e., the pairwise correlation matrix $\Gamma_{ab}$), the residuals are assumed to follow the standard multivariate Gaussian model as in existing analyses. At the next level of the hierarchy, the ORF itself is assigned a non-Gaussian prior whose mean is anchored to the HD curve, while its fluctuations are informed by the cosmic variance and cosmic skewness (and, if needed, by higher-order cumulants). Physically, this reflects the fact that a GWB is a stochastic process: even for an isotropic population, the “true” ORF realized in our Universe may deviate from its ensemble mean (HD) due to finite-source and interference effects. The hierarchical construction therefore incorporates not only measurement uncertainty but also the uncertainty associated with the cosmic realization of the background. In this sense, our proposal can be viewed as a complementary extension that aims to incorporate such cosmic-realization uncertainty directly within the residual-level Gaussian-likelihood framework used in standard PTA pipelines.

At the level of a schematic model, the likelihood can be written as
\begin{equation}
\mathcal{L}\ \propto\ p(\delta t \mid \Gamma, A_{\rm GW}, \ldots)\,p(\Gamma \mid N)\,p(N),
\end{equation}
where the hyperparameter $N$ represents an effective number of sources contributing within a given frequency bin (more generally, a frequency-dependent $N(f)$). This approach has several potential advantages over conventional analyses. First, while preserving the HD correlation as the ensemble mean, it provides a principled way to account for astrophysically expected deviations induced by source discreteness, thereby reducing the risk of misinterpreting modest departures from HD as evidence for noise systematics or non-GW processes. Second, insofar as the data retain information about ORF fluctuations (cosmic variance) and asymmetry (cosmic skewness), the framework opens a route to constraining the effective source number through hyperparameter inference. In fact, as shown in Fig.~\ref{fig:cosmic_skew_amplitude}, the cosmic skewness increases with $N$ for small $N$, while in the large-$N$ limit ($N > 100$) it tends to saturate. Although $N$ itself may be difficult to identify especially for large $N$, constraints are more likely to manifest as bounds on an effective discreteness measure rather than a precise determination of $N$.

Finally, as suggested by our results, the normalized skewness is typically of order unity, implying that third-order moments alone may not fully capture the behavior of the distribution tails (rare outliers). For practical applications, skewness should be viewed as a first step toward characterizing non-Gaussianity. Higher-order statistics (e.g., kurtosis and beyond)~\cite{2025JCAP...01..017B}, cumulant-based approximations (such as Edgeworth expansions), or direct likelihood modeling of the relevant estimators may be required for stronger discriminating power, particularly in regimes dominated by a small number of bright sources. These will be pursued in future work.

\acknowledgements
KT is partially supported by JSPS KAKENHI Grant Numbers 20H00180, 21H01130, 21H04467, 24H01813 and 25K21670 and Bilateral Joint Research Projects of JSPS 120237710.

\appendix
\renewcommand{\theequation}{\thesection.\arabic{equation}}
\section{\label{app:many}Many-point Skewness}
In this appendix, we provide the detailed derivation of the dominant term for the third moment in the interfering model. The integrand is defined in Eq. (\ref{eq:rho^3}). After performing the phase-averaging process, Eq. (\ref{eq:phase averaged rho^3}) yields three non-vanishing terms.
We note that we describe everything except the directional average of non-dominant terms.

\begin{widetext}
\begin{enumerate}
    \item The \(\langle\rho_\text{diag}^3\rangle\) Term

  As shown in Sec. \ref{sec:many}, the \(\langle\rho_\text{diag}^3\rangle\) term is independent of phase. The expansion contains terms proportional to \(\mathcal{H}_6\), \(\mathcal{H}_4 \mathcal{H}_2\), and \(\mathcal{H}_2^3\). In the large-\(N\) limit, the contribution from the \(i \neq j\neq k\) terms becomes dominant. Consequently,  the ensemble average yields:
  \begin{align}\label{eq-dom-rho^3-diag}
   \langle\rho_\text{diag}^3\rangle 
    &=
    \left\langle
    \biggl(
     \sum_i (c_i d_i^* + c_i^* d_i)
    \biggl)^3
    \right\rangle_\Omega \notag\\
    &=
    \left\langle
     \sum_i (c_i d_i^* + c_i^* d_i)^3
    \right\rangle _\Omega
    +3
    \left\langle
    \sum_{i\neq j}(c_i d_i^* + c_i^* d_i)(c_j d_j^* + c_j^* d_j)^2
    \right\rangle_\Omega\nonumber\\
&\quad+\left\langle\sum_{i\neq j\neq k}(c_i d_i^* + c_i^* d_i)(c_j d_j^* + c_j^* d_j)(c_k d_k^* + c_k^* d_k)
    \right\rangle_\Omega\notag\\
    &\approx\frac{1}{8}\mathcal{H}_2^3\mu_\text{u}^3(\gamma),
  \end{align}
where $\left\langle\sum_i(c_i d_i^* + c_i^* d_i)^3\right\rangle$ is proportional to $\mathcal{H}_6$. The term $3\left\langle\sum_{i\neq j}(c_i d_i^* + c_i^* d_i)(c_j d_j^* + c_j^* d_j)^2\right\rangle$ accounts for the three cases arising from the cyclic permutations of $i=j\neq k$, and it proportional to \(\mathcal{H}_4 \mathcal{H}_2\), thus the dominant term proportional to the cube of HD curve.
 
    \item The \(\langle3\rho_\text{diag}\rho_\text{off-diag}^2\rangle\) Term

\begin{align}\label{eq:phase averaged 3rho-diag rho-offdiag^2}
  & 
  \left\langle
   3\rho_\text{diag}\rho_\text{off-diag}^2
  \right\rangle\notag\\
  &=
  \Biggl\langle3 \sum_i (c_i d_i^* + c_i^* d_i)
   \Biggl[ \sum_{j\neq k}
    \left\{
     c_j d_k^* e^{i(\phi_j - \phi_k)} + c_j^* d_k e^{-i(\phi_j - \phi_k)}
    \right\}
    \Biggl]
   \Biggl[ \sum_{\ell\neq m}
    \left\{
     c_j d_k^* e^{i(\phi_\ell - \phi_m)} + c_\ell^* d_m e^{-i(\phi_\ell - \phi_m)}
    \right\}
   \Biggl]
   \Biggl\rangle\notag\\
    &=
  \Biggl \langle 3 \sum_i (c_i d_i^* + c_i^* d_i)
   \sum_{j\neq k} \sum_{\ell\neq m}
   \Bigl[ 
    c_j d_k^* c_\ell d_m^* 
    \langle 
     e^{i(\phi_j-\phi_k+\phi_\ell-\phi_m)} 
    \rangle_\phi +
    c_j d_k^* c_\ell^* d_m 
    \langle 
     e^{i(\phi_j-\phi_k-\phi_\ell+\phi_m)} 
    \rangle_\phi +\notag\\
   &c_j^* d_k c_\ell^* d_m 
    \langle 
     e^{-i(\phi_j-\phi_k+\phi_\ell-\phi_m)} 
    \rangle_\phi +
    c_j^* d_k c_\ell d_m^* 
    \langle 
     e^{-i(\phi_j-\phi_k-\phi_\ell+\phi_m)} 
    \rangle_\phi 
   \Bigl] 
  \Biggl\rangle.
\end{align}
\begin{enumerate}
     \item Phase Averaging:
     
     The phase average \(\langle...\rangle_\phi\) is performed on the product of the four off-diagonal phase factors. The two contributing terms, \(c_j d_k^* c_\ell d_m^* \langle e^{i(\phi_j-\phi_k+\phi_\ell-\phi_m)} \rangle_\phi\) and \(c_j d_k^* c_\ell^* d_m \langle e^{i(\phi_j-\phi_k-\phi_\ell+\phi_m)} \rangle_\phi\), lead to combinations of Kronecker deltas (\(\delta_{jm}\delta_{k\ell}\) and \(\delta_{j\ell}\delta_{km}\)) that enforce index pairings. The resulting expression, before direction averaging is given by:
  \begin{align}\label{eq:completed phase averaged 3rho-diag rho-offdiag^2}
   \left\langle
    3\rho_\text{diag}\rho_\text{off-diag}^2
   \right\rangle_\phi
    &=
   \Biggl (
    3\sum_i (c_i d_i^* + c_i^* d_i)
   \Biggl )
   \sum_{j\neq k}
   \Bigl[ 
    c_j d_k^* c_k d_j^*  
  + c_j d_k^* c_j^* d_k   
  + c_j^* d_k c_k^* d_j 
  + c_j^* d_k c_j d_k^*  
   \Bigl] .
  \end{align}
    \item Direction Averaging:
  
  The direction average \(\langle...\rangle_\Omega\) contains three indices (\(i,j,k\)). We can divide Eq. (\ref{eq:completed phase averaged 3rho-diag rho-offdiag^2}) into two parts: \(6\sum_{i=j\neq k}\) and \(3\sum_{i\neq j\neq k}\). The dominant term is shown as follows:
  \begin{align}\label{eq-3rho-diag-rho^2-off-diag}
  \left\langle
   3\rho_\text{diag}\rho_\text{off-diag}^2
  \right\rangle
 &=6
  \left\langle
  \sum_{j\ne k}(c_jd_j^*+c_j^*d_j)
  (c_j d_k^* c_k d_j^*  
  + c_j d_k^* c_j^* d_k   
  + c_j^* d_k c_k^* d_j 
  + c_j^* d_k c_j d_k^*  )
  \right\rangle_\Omega\notag\\
   &\quad+3\sum_{i\neq j\neq k} 
  \langle
   (c_i d_i^* + c_i^* d_i)
  \rangle_\Omega
  \bigl[
   \langle
    c_j d_j^*
   \rangle_\Omega
   \langle
    c_k d_k^*
   \rangle_\Omega
   +
   \langle
    c_j^* d_j
   \rangle_\Omega
   \langle
    c_k^* d_k
   \rangle_\Omega
   +2 
   \langle
    |c_j|^2
   \rangle_\Omega 
   \langle
    |d_k|^2
   \rangle_\Omega 
  \bigl]\notag\\
 &\approx\frac{3}{16}
  \mathcal{H}_2^3
  \Bigl[
   \mu_\text{u}^3(\gamma) 
+  \mu_\text{u}(\gamma)(1+\chi^2)^2\mu_\text{u}^2(0)
  \Bigl].
  \end{align}
  
  \end{enumerate}

\item The \(\langle\rho_\text{off-diag}^3\rangle\) Term

  This term represents the cubic self-interaction of the off-diagonal interference terms, written as:
  \begin{align}
  &\langle
    \rho_\text{off-diag}^3
   \rangle_\phi\nonumber\\
   =&
   \Biggl\langle\sum_{i\neq j}\sum_{k\neq \ell}\sum_{m\neq n} 
    \Bigl[ 
     c_i d_j^* e^{i(\phi_i - \phi_j)} 
   + c_i^* d_j e^{-i(\phi_i - \phi_j)}
    \Bigl]
    \Bigl[
     c_k d_\ell^* e^{i(\phi_k - \phi_\ell)} 
   + c_k^* d_\ell e^{-i(\phi_k - \phi_\ell)}
    \Bigl]\nonumber\\
    \times
   &\Bigl[
     c_m d_n^* e^{i(\phi_m - \phi_n)} 
   + c_m^* d_n e^{-i(\phi_m - \phi_n)}
    \Bigl]
  \Biggl\rangle_\phi\nonumber\\
  =&
  \sum_{i\neq j}\sum_{k\neq \ell}\sum_{m\neq n}
  \Bigl[ 
   c_i d_j^* c_k d_\ell^* c_m d_n^* 
   \langle 
    e^{i(\phi_i - \phi_j + \phi_k - \phi_\ell + \phi_m - \phi_n)} 
   \rangle_\phi 
   +c_i^* d_j c_k^* d_\ell c_m^* d_n 
   \langle 
    e^{-i(\phi_i - \phi_j + \phi_k - \phi_\ell + \phi_m - \phi_n)} 
   \rangle_\phi\nonumber\\
& + c_i d_j^* c_k d_\ell^* c_m^* d_n 
   \langle 
    e^{i(\phi_i - \phi_j - \phi_k + \phi_\ell + \phi_m - \phi_n)} 
   \rangle_\phi 
  + c_j^* d_j c_k^* d_\ell c_m d_n ^*
   \langle 
    e^{-i(\phi_i - \phi_j - \phi_k + \phi_\ell + \phi_m - \phi_n)} 
   \rangle_\phi\nonumber\\
& + c_i d_j^* c_k^* d_\ell c_m d_n^* 
   \langle 
    e^{i(\phi_i - \phi_j - \phi_k + \phi_\ell + \phi_m - \phi_n)} 
   \rangle_\phi 
  + c_i^* d_j c_k d_\ell^* c_m^* d_n 
   \langle e^{-i(\phi_i - \phi_j - \phi_k + \phi_l + \phi_m - \phi_n)} 
   \rangle_\phi\nonumber\\
& + c_i d_j^* c_k^* d_\ell c_m ^*d_n 
   \langle 
    e^{i(\phi_i - \phi_j - \phi_k + \phi_\ell - \phi_m + \phi_n)} 
   \rangle_\phi
  + c_i^* d_j c_k d_\ell^* c_m d_n^* 
   \langle 
    e^{-i(\phi_i - \phi_j - \phi_k + \phi_\ell - \phi_m + \phi_n)} 
   \rangle_\phi
  \Bigl].
  \end{align}
  
  \begin{enumerate}
   \item Phase Averaging

  The phase average \(\langle...\rangle_\phi\) is applied to the expansion of the cubed off-diagonal sum, which involves a summation over six indices (\(i\neq j,k\neq \ell,m\neq n\)). The calculation requires evaluating six-phase correlation, such as \(\langle e^{i(\phi_i - \phi_j + \phi_k - \phi_\ell + \phi_m - \phi_n)} \rangle_\phi\). These averages are non-zero only when the six indices can be completely paired off to eliminate the phase dependence. This results in a sum of terms involving products of three Kronecker deltas. For example, the specific phase combinations yield:
\begin{align}\label{delta_combination}
 \langle 
  e^{i(\phi_i - \phi_j + \phi_k - \phi_\ell + \phi_m - \phi_n)}
 \rangle_\phi 
 &= \delta_{i\ell} \delta_{jm} \delta_{kn} 
 + \delta_{in} \delta_{jk} \delta_{\ell m} \nonumber\\
 \langle 
  e^{i(\phi_i - \phi_j + \phi_k - \phi_\ell - \phi_m + \phi_n)} 
 \rangle_\phi 
 &= \delta_{i\ell} \delta_{jn} \delta_{km} 
 + \delta_{im} \delta_{jk} \delta_{\ell n} \nonumber\\
 \langle 
  e^{i(\phi_i - \phi_j - \phi_k + \phi_\ell + \phi_m - \phi_n)} 
 \rangle_\phi 
 &= \delta_{ik} \delta_{jm} \delta_{\ell n} 
 + \delta_{in} \delta_{j\ell} \delta_{km} \nonumber\\
 \langle 
  e^{i(\phi_i - \phi_j - \phi_k + \phi_\ell - \phi_m + \phi_n)} 
 \rangle_\phi 
 &= \delta_{ik} \delta_{jn} \delta_{\ell m} 
 + \delta_{im} \delta_{j\ell} \delta_{kn}.
\end{align}
Applying these delta constraints to the full expansion reduces the six-fold sum to a three-fold sum, where the terms consist of products of coefficients \(c\) and \(d\).

  \item Direction Averaging:

  The resulting expression after phase averaging is then averaged over source directions \(\langle...\rangle_\Omega\). Similar to the previous terms, this yields contributions of varying orders of \(\mathcal{H}\). We again  focus on the dominant contribution where the three source pairs are distinct (\(i\neq j \neq k\)). In this limit, the statistical independence allows the averages to factorize into products of first and second moments.

  The dominant term expands into a sum of four distinct types of moment products:
\begin{align}\label{eq-rho^3-off-diag}
 \langle
   \rho_\text{off-diag}^3
  \rangle
 &=\,4\sum_{i\neq j\neq k}
 \Bigl[
  \langle c_i d_i^* \rangle_\Omega
  \langle c_j d_j^* \rangle_\Omega
  \langle c_k d_k^* \rangle_\Omega
  +
3 \langle c_i d_i^* \rangle_\Omega 
  \langle |c_k|^2 \rangle_\Omega
  \langle |d_j|^2 \rangle_\Omega 
 \Bigl]\nonumber\\
 &\approx
 \frac{1}{16} \mathcal{H}_2^3
 \Bigl[
  \mu_\text{u}^3(\gamma) 
+3\mu_\text{u}(\gamma)(1+\chi^2)^2\mu_\text{u}^2(0)
 \Bigl].
\end{align}
  \end{enumerate}
  
  \end{enumerate}
Finally, by combining Eq.~\eqref{eq-dom-rho^3-diag}, Eq.~\eqref{eq-3rho-diag-rho^2-off-diag}, and Eq.~\eqref{eq-rho^3-off-diag}, we obtain the dominant third moment Eq.~\eqref{eq:dom rho^3}

\section{\label{app:cosmic}Three-point-average function}

In this Appendix, we detail the calculation of the three-point function \(\widehat{\mu^3}(\gamma)\), which is necessary to evaluate the cosmic skewness derived in Sec.\ref{sub:cosmic skew}.
We begin with the analytic form of the two-point function \(\mu(\gamma,\beta)\) derived by Allen (2023)\cite{Allen2023_Variance}. This function describes the correlation between two pulsars separated by an angle \(\gamma\) due to interference from two GW sources separated by an angle \(\beta\). The function is defined piecewise as:

\begin{align}\label{eq-mu-gamma-beta-explicit}
\mu(\gamma, \beta) = 
\begin{cases}
\begin{aligned}
&
\frac{1}{48}
 \left[
  33-18 \cos \beta-3 \cos ^2 \beta+
  \left(
   32-21 \cos \beta-6 \cos ^2 \beta-\cos ^3 \beta
  \right) 
  \cos \gamma
 \right] 
  \sec ^4
  \left(
   \frac{\beta}{2}
  \right) \\
&+
 \left(
  1-\frac{1}{2} \cos \beta-\frac{1}{2} \cos \gamma
 \right) 
 \sec ^4
 \left(
  \frac{\beta}{2}
 \right)
 \log 
 \left(
  \sin ^2
  \left(
   \frac{\gamma}{2}
  \right)
 \right)
\end{aligned} & \text{for } \beta < \gamma \\
\\
\begin{aligned}
&
 \frac{1}{24}
 \left[
  33-3 \cos \beta-
  \left(
   16+5 \cos \beta+\cos ^2 \beta
  \right) 
  \cos \gamma
 \right] 
 \sec ^2
 \left(
  \frac{\beta}{2}
 \right) \\
&+
 \left(
  1-\frac{1}{2} \cos \beta-\frac{1}{2} \cos \gamma
 \right) 
 \sec ^4
 \left(
  \frac{\beta}{2}
 \right) 
 \log 
 \left(
  \sin ^2
  \left(
   \frac{\beta}{2}
  \right)
 \right)
\end{aligned} & \text{for } \beta > \gamma
\end{cases}
\end{align}
    
\end{widetext}
\begin{figure}
  \centering
  \includegraphics[width=0.45\textwidth]{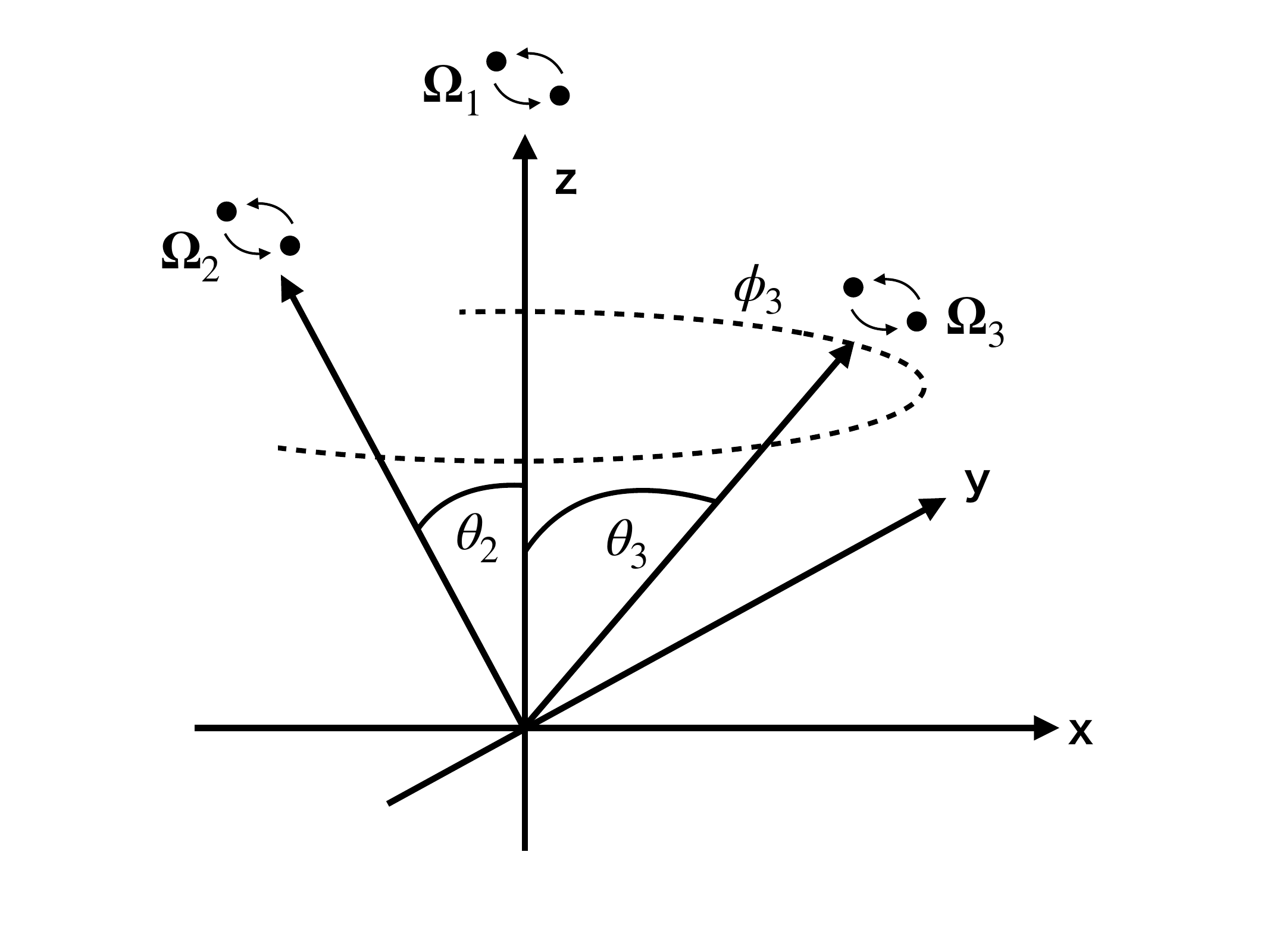}
  \caption{The direction of the first GW source is along the $\boldsymbol{}{z}$ axis. The direction of the second GW source exists in the $xz$-plane parameterized by ($\theta_2$). The third GW source takes any direction by ($\theta_3,\phi_3$).}
  \label{fig:cosmic_set}
\end{figure}
For notational brevity in the subsequent integration, we define the functions \(A(\gamma, \beta)\) and \(B(\gamma, \beta)\) corresponding to the two cases in Eq. (\ref{eq-mu-gamma-beta-explicit}):
\begin{align}
\mu(\gamma, \beta) = 
\begin{cases}
A(\gamma,\beta) & \text{for } \beta < \gamma \\
B(\gamma,\beta) & \text{for } \beta > \gamma
\end{cases}
\end{align}
The calculation of the three-point function \(\widehat{\mu^3}(\gamma) = \langle \mu(\gamma, \beta_{12})\mu(\gamma, \beta_{13})\mu(\gamma, \beta_{23}) \rangle\) requires averaging over the configurations of three independent GW sources. To parameterize this average, we define the angular positions of the three sources \(\boldsymbol{\Omega}_1, \boldsymbol{\Omega}_2, \boldsymbol{\Omega}_3\) using two polar angles (\(\theta_2, \theta_3\)) and one azimuthal angle (\(\phi_3\)).
Without loss of generality, we align the first source with the \(z\)-axis and place the second source in the \(xz\)-plane:
\begin{align}
  \boldsymbol{\Omega}_1 &= \boldsymbol{z} \\
  \boldsymbol{\Omega}_2 &= \boldsymbol{x}\sin\theta_2  + \boldsymbol{z} \cos\theta_2 \\
  \boldsymbol{\Omega}_3 &= \boldsymbol{x}\sin\theta_3\cos\phi_3  + \boldsymbol{y} \sin\theta_3\sin\phi_3 + \boldsymbol{z}\cos\theta_3
\end{align}
Based on these coordinates, the separation angles \(\beta_{mn}\) between sources \(m\) and \(n\) are given by:
\begin{align}
  \cos\beta_{12} &= \cos\theta_2 \\
  \cos\beta_{13} &= \cos\theta_3 \\
  \cos\beta_{23} 
 &= \boldsymbol{\Omega}_2 \cdot \boldsymbol{\Omega}_3\notag\\
 &= \sin\theta_2\sin\theta_3\cos\phi_3 + \cos\theta_2\cos\theta_3
\end{align}
The ensemble average is computed by integrating over the solid angles of the sources. Taking into account the rotational symmetry around \(\boldsymbol{\Omega}_1\) (which integrates out \(\phi_2\) to \(2\pi\)) and normalizing by the total solid angle phase space \((4\pi)^2\), the expression for the three-point function becomes:
\begin{align}\label{eq-mu3-integral}
 \widehat{\mu^3}(\gamma)
 =& 
 \frac{1}{8\pi} 
 \int_0^\pi d\theta_2 \sin\theta_2 
 \int_0^\pi d\theta_3 \sin\theta_3 
 \int_0^{2\pi} d\phi_3\notag\\
 &\times
 \mu(\gamma, \theta_2) 
 \mu(\gamma, \theta_3) 
 \mu(\gamma, \beta_{23}(\theta_2, \theta_3, \phi_3))
\end{align}
The integrands for \(\beta_{12}\) and \(\beta_{13}\) are straightforward, as \(\theta_2\) and \(\theta_3\) directly correspond to the integration variables. We can split the integrals based on the domains \(\theta < \gamma\) and \(\theta > \gamma\).
However, the integration over \(\phi_3\) presents a significant challenge. The function \(\mu(\gamma, \beta_{23})\) switches between the forms \(A(\gamma, \beta_{23})\) and \(B(\gamma, \beta_{23})\) depending on whether \(\beta_{23} < \gamma\). The condition \(\beta_{23}(\theta_2, \theta_3, \phi_3) < \gamma\) defines a complex boundary in the \((\theta_2, \theta_3, \phi_3)\) configuration space:
\begin{align}
 \sin\theta_2\sin\theta_3\cos\phi_3 + \cos\theta_2\cos\theta_3 > \cos\gamma
\end{align}
Because of the complex functional dependence of \(\beta_{23}\) on the integration variables, coupled with the intricate form of the functions \(A\) and \(B\) (which contain logarithmic and secant terms), an analytical evaluation of this integral is intractable. Therefore, in this study, we perform the integration of Eq. (\ref{eq-mu3-integral}) numerically to obtain the values of \(\widehat{\mu^3}(\gamma)\).


\bibliography{main}    

\end{document}